\documentclass[10pt, a4paper]{article}
\usepackage{jheparxiv}
\usepackage[utf8]{inputenc}
\usepackage{amsmath}
\usepackage{amsmath,bm}

\usepackage{amsfonts}\usepackage{amssymb}
\usepackage{latexsym}
\usepackage{mathrsfs}
\usepackage{braket}		
\usepackage{graphicx}
\usepackage{color}
\usepackage{xcolor}
\usepackage{slashed}
\usepackage{mathtools}
\usepackage{breqn}
\usepackage[all]{xy}
\usepackage{subcaption}

\newcommand{\RNum}[1]{\uppercase\expandafter{\romannumeral #1\relax}}

\subheader{\hfill \texttt{}}

\definecolor{hyperref}{RGB}{026,028,087}

\newcommand{\beq}{\begin{equation}}
\newcommand{\eeq}{\end{equation}}
\newcommand{\bea}{\begin{eqnarray}}
\newcommand{\eea}{\end{eqnarray}}
\def\be{\begin{equation}}
\def\ee{\end{equation}}

\newcommand{\mpl}{M_{\rm Pl}}

\renewcommand{\L}{\mathcal L}

\renewcommand{\O}{\mathcal O}

\newcommand{\del}{\partial}
\newcommand{\ri}{i}

\newcommand{\rd}{\mathrm{d}}
\newcommand{\cO}{\mathcal{O}}

\newcommand{\msbar}{\overline{\text{MS}}}
\newcommand{\ddp}{\frac{\rd^{d}p}{(2\pi)^d}}
\newcommand{\ddl}{\frac{\rd^{d}l}{(2\pi)^d}}

\newcommand{\kep}{(k\varepsilon)}
\newcommand{\vep}{\varepsilon}
\newcommand{\fa}{\mathfrak{a}
}
\newcommand{\fb}{\mathfrak{b}}
\newcommand{\fc}{\mathfrak{c}}
\newcommand{\fd}{\mathfrak{d}}

\newcommand{\disc}{\mathrm{Disc}}
\newcommand{\cA}{\mathcal{A}}

\newcommand{\eref}{\eqref}

\def\be{\begin{equation}}
\def\ee{\end{equation}}
\def\ba{\begin{eqnarray}}
\def\ea{\end{eqnarray}}

\def\p{\partial}
\def\d{\mathrm{d}}
\def\mn{_{\mu \nu}}

\def\({\left(}
\def\){\right)}

\def\nn{\nonumber}

\def\d{\mathrm{d}}

\usepackage[normalem]{ulem}

\def\ba{\begin{eqnarray}}
\def\ea{\end{eqnarray}}

\begin{document}

\title{QED positivity bounds}

\author[a]{Lasma Alberte,}
\author[a,b]{Claudia de Rham,}
\author[a]{Sumer Jaitly,}
\author[a,b]{Andrew J. Tolley}
\affiliation[a]{Theoretical Physics, Blackett Laboratory, Imperial College, London, SW7 2AZ, U.K.}
\affiliation[b]{CERCA, Department of Physics, Case Western Reserve University, 10900 Euclid Ave, Cleveland, OH 44106, USA}

\emailAdd{l.alberte@imperial.ac.uk}
\emailAdd{c.de-rham@imperial.ac.uk}
\emailAdd{sumer.jaitly14@imperial.ac.uk}
\emailAdd{a.tolley@imperial.ac.uk}

\abstract{
We apply positivity bounds directly to a $U(1)$ gauge theory with charged scalars and charged fermions, i.e. QED, minimally coupled to gravity. Assuming that the massless $t$-channel pole may be discarded, we show that the improved positivity bounds are violated unless new physics is introduced at the parametrically low scale $\Lambda_{\rm new} \sim (e m \mpl)^{1/2}$, consistent with similar results for scalar field theories, far lower than the scale implied by the weak gravity conjecture. This is sharply contrasted with previous treatments which focus on the application of positivity bounds to the low energy gravitational Euler-Heisenberg effective theory only. We emphasise that the low-cutoff is a consequence of applying the positivity bounds under the assumption that the pole may be discarded. We conjecture an alternative resolution that a small amount of negativity, consistent with decoupling limits,  is allowed and not in conflict with standard UV completions, including weakly coupled ones.}

\maketitle


\section{Introduction}

It is now well established that for non-gravitational quantum field theories to admit a local Lorentz invariant unitary UV completion, the low energy scattering amplitude should satisfy an array of positivity bounds that constrain the sign and magnitude of Wilson coefficients.  The simplest bounds were first noted in \cite{Pham:1985cr,Ananthanarayan:1994hf,Adams:2006sv} and the connection between their violation and causality was emphasized in \cite{Adams:2006sv}. These original forward limit scalar bounds have been extended to general spins \cite{Bellazzini:2016xrt,deRham:2017zjm} away from the forward limit \cite{deRham:2017avq,deRham:2017zjm}. These bounds have proven fruitful in placing constraints on interacting spin-2 fields \cite{Cheung:2016yqr,Bonifacio:2016wcb,deRham:2017imi,deRham:2017xox,deRham:2018qqo,Alberte:2019xfh,Alberte:2019zhd,Wang:2020xlt}, restricting beyond standard model interactions \cite{Zhang:2018shp,Bi:2019phv,Remmen:2019cyz,Zhang:2020jyn,Remmen:2020uze,Remmen:2020vts,Yamashita:2020gtt,Trott:2020ebl,Bonnefoy:2020yee,Fuks:2020ujk}, and providing a new light on properties of string amplitudes \cite{Green:2019tpt,Huang:2020nqy}. Most recently it has been recognized that by using more information from crossing symmetry and the partial wave expansion it is possible to put upper and lower bounds on Wilson coefficients \cite{Tolley:2020gtv,Bellazzini:2020cot,Caron-Huot:2020cmc,Sinha:2020win,AHH} in certain cases ruling out classes of theories from having a standard UV completion such as weakly broken Galileon theories \cite{Bellazzini:2020cot,Tolley:2020gtv}. Similar results are arrived at within the related S-matrix bootstrap program \cite{Guerrieri:2020bto}. \\

Given these successes, it is highly desirable to consider the impact of these bounds for realistic effective field theories coupled to gravity. Unfortunately, the direct application of positivity bounds to gravitational effective field theories is fraught with difficulties. On the one hand the distinctive features of gravity mean that scattering amplitudes are permeated by massless poles and branch points which spoil the conventional forward limit considerations, and preclude an analytic continuation from the physical region which preserves positivity even away from the forward limit. On the other hand, causality in the gravitational setting is more subtle, from the ambiguity of the metric under field redefinitions and known superluminal speeds in well established low energy effective field theories (EFTs) \cite{Drummond:1979pp,Hollowood:2015elj,deRham:2019ctd,deRham:2020zyh}. In a previous paper \cite{Alberte:2020jsk}, we argued that the only gravitational effective theories in which positivity is clear cut are those for which there is a clean $\mpl \rightarrow \infty$ decoupling limit, for which positivity of the non-gravitational decoupling limit theory may be assured. With this in mind, we considered several examples of renormalizable scalar field theories coupled to gravity for which violations of positivity are necessarily suppressed by powers of $\mpl$. Demanding the scattering amplitude respects positivity with the gravitational $t$-channel pole removed generically imposes the cutoff of the effective theory to be far lower than expected, a result which parallels conclusions from the swampland program \cite{Vafa:2005ui,Ooguri:2006in}. \\

In the present work we extend the results of  \cite{Alberte:2020jsk} to the more interesting case of QED minimally coupled to gravity. Unlike in Ref.~\cite{Alberte:2020jsk}, we will not rely on the device of introducing a spectator field, but rather consider the improved positivity bounds \cite{deRham:2017xox,deRham:2017imi,Bellazzini:2016xrt}. In their simplest form, the standard forward limit positivity bounds can be applied on the pole-subtracted scattering amplitudes as \cite{Adams:2006sv}:
\be\label{positivity}
\frac{\text{d}^2\mathcal A_i(s,0)}{\text{d}s^2}=\frac{2}{\pi} \int_{4m^2}^{\infty} \d \mu \frac{\text{Im}\,\mathcal A_i(\mu,0)}{(\mu-s)^3} +\frac{2}{\pi} \int_{4m^2}^{\infty}  \d \mu \frac{\text{Im}\,\mathcal A^{\rm c}_i(\mu,0)}{(\mu-u)^3} >0\,,
\ee
where the positivity of the expression (for $ 0<s< 4m^2$, $u=4m^2-s$) on the left-hand side arises due to the analyticity properties of the S-matrix and positivity from the optical theorem, and $A^{\rm c}_i$ is the $s$-$u$ crossing exchanged amplitude. The improved positivity bounds \cite{deRham:2017xox,deRham:2017imi,Bellazzini:2016xrt} allow us to tighten the bound by including any additional knowledge about our EFT. The idea behind them is particularly transparent from the exact formulation of the optical theorem as:
\be\label{optical}
\text{Im}\,\mathcal A_i(s,0)=\frac{1}{2}\sum_f\int\text{d}\Pi_f |\mathcal A_{i\to f}|^2>0\,,
\ee
where $i$ denotes the initial and final particle content, $f$ stands for any intermediate state and $\text{d}\Pi_f$ is the phase space volume. The theorem then implies that, given a set of possible intermediate states in the theory that is being investigated, i.e. $\{f_1,f_2,\dots,f_N\}$, each known contribution to the sum on the right-hand side of the above equation can be taken to the left-hand side leading to an even tighter constraint on the remaining amplitudes. This gives the improved positivity bounds
\be\label{improvedpositivity}
\frac{\text{d}^2\mathcal A_i(s,0)}{\text{d}s^2}-\frac{1}{\pi} \sum_{\text{known }f}\int\text{d}\Pi_f   \int_{4m^2}^{\infty}  \d \mu  \frac{ |\mathcal A_{i\to f}|^2}{(\mu-s)^3} -\frac{1}{\pi} \sum_{\text{known }f}\int\text{d}\Pi_f   \int_{4m^2}^{\infty}  \d \mu  \frac{ |\mathcal A^c_{i\to f}|^2}{(\mu-u)^3} >0\,,
\ee
where overall positivity is still ensured by the sum over `unknown' configurations $f$. It is in the application of improved positivity bounds that our results will differ from previous discussions of positivity bound for QED coupled to gravity, notably \cite{Cheung:2014ega}, and more recently \cite{Hamada:2018dde,Bellazzini:2019xts,Chen:2019qvr} which have focused entirely on the gravitational Euler-Heisenberg effective field theory that describes physics well below the electron mass\footnote{This information is partly recovered in the 3D case considered in \cite{Chen:2019qvr} by focussing on the large order limit in an expansion in $s/m^2$. In practice, for our considerations it is better to utilize the improved positivity bounds since the former is dominated by the branch put at $4m^2$ and the latter at a much higher scale.}. The latter is sufficient to reproduce the bounds \eqref{positivity}, but by preserving information from physics at and above the electron mass, one is able to derive a much tighter constraint as implied by the improved bound \eqref{improvedpositivity}.  \\

Remarkably, the authors of \cite{Cheung:2014ega} noted that if positivity bounds were applied to 4-photon (i.e. $2$-$2$) scattering amplitudes with the gravitational $t$-channel exchange removed\footnote{These bounds can be motivated on entropic grounds \cite{Cheung:2018cwt,Loges:2019jzs,Cheung:2019cwi,Goon:2019faz} or in other setups \cite{Andriolo:2018lvp,Aalsma:2019ryi,Cremonini:2019wdk,Aalsma:2020duv}. Recently, the procedure of applying directly the positivity bounds to the $t$-channel removed amplitude was argued to be justified by a compactification argument in \cite{Bellazzini:2019xts}. In \cite{Alberte:2020jsk} various issues with this compactification argument were pointed out. See also \cite{Tokuda:2020mlf,Herrero-Valea:2020wxz} for related discussions.}, positivity would hold if the general requirements of the weak gravity conjecture  \cite{ArkaniHamed:2006dz} are met, namely that there is a bound on the charge to mass ratio $|e|/m\gtrsim 1/\mpl$.
Interestingly, at least in 3D, this observation is partly countered by that of \cite{Chen:2019qvr} which uses the extended positivity bounds of \cite{AHH} to derive opposing bounds, arguing for the need for additional light neutral states to resolve this tension. As we discuss in section~\ref{higherorder}, this particular `resolution' does not apply in the four dimensional case considered here. \\

Keeping in the spirit of applying positivity bounds to the $t$-channel removed amplitude, we shall find a much stronger result: {\it Improved positivity bounds applied to QED coupled to gravity demand the existence of new physics at the scale $\Lambda_{\rm new}\sim (e m \mpl)^{1/2}$}. Most importantly this result is independent of what that new physics is. For instance, it applies equally well for the Regge like completions considered in \cite{Hamada:2018dde} where the photon Regge tower dominates over the graviton tower, and it is argued that the weak gravity conjecture from positivity argument is robust. That is because any Lorentz invariant UV completion will be described at low energies as irrelevant operators correcting the naive QED Lagrangian, and our consideration only demands that some new physics comes in at the scale $\Lambda_{\rm new}\sim (e m \mpl)^{1/2}$, which would show up at low energies as the need to add irrelevant operators, but makes no demands to what its origin is. \\

As discussed in \cite{Alberte:2020jsk} an alternative explanation of our results is that strict positivity of the scattering amplitude, with the $t$-channel pole removed, does not apply. Indeed we can only be sure it applies in the decoupling limit $\mpl \rightarrow \infty$. Rather in \cite{Alberte:2020jsk} we conjectured that in the gravitational context, for a scattering amplitude whose low energy expansion near $t=0$ takes the form\footnote{In general graviton loops lead to branch cuts extending to $t=0$, however for the 4-photon amplitude these necessarily arise at order $1/\mpl^4$ and so will not affect any considerations here. Nevertheless, they are indicative of the issues with continuing the partial wave expansion past $t=0$.}
\be
{\cal A} \sim -\frac{s^2}{\mpl^2 t}  + c \,  s^2  + \dots
\ee
the standard positivity bound \eqref{positivity} is weakened to the requirement
\be \label{newpositivitiy}
c >- \frac{{\cal O}(1)}{M^2\mpl^2}  \, ,
\ee
where $M$ is at most the cutoff $\Lambda_c$ of the low energy expansion $M\le \Lambda_c$. This weakening is consistent with the known weakening of causality criteria in familiar EFTs \cite{Drummond:1979pp,deRham:2019ctd,deRham:2020zyh,Hollowood:2015elj}. Our results for QED indicate that the improved positivity bound \eqref{improvedpositivity} would need to be weakened to
\be\label{newimprovedpositivity}
c^{\rm imp} >- \frac{e^2}{m^2\mpl^2} \times {\cal O}(1) \, ,
\ee
where $m$ is the electron mass to avoid the need to introduce new physics at the scale $\Lambda_{\rm new}\sim (e m \mpl)^{1/2}$. Here $c^{\rm imp} $ is the equivalent coefficient that arises in the expansion of the improved amplitude \eqref{improvedamplitude}. This is consistent with \eqref{newpositivitiy} for $M \sim m/e$. While \eqref{newimprovedpositivity} is not in conflict with the $\mpl \rightarrow \infty$ decoupling limit, it would nevertheless indicate a significant weakening of positivity that deserves further explanation. At present their is no generally accepted proof of positivity of $c$ at finite $\mpl$.\\

We stress again that our conclusions are valid for generic standard UV completions and further assuming weak coupling, by itself, would not improve the bound \eqref{newimprovedpositivity}. The UV completion may equally well be strongly coupled at the scale $\Lambda_{\rm new}$ or lead to an infinite tower of higher spin states as is required in any tree level completion of gravity such as string theory. We only require that QED minimally coupled to gravity be a good description at low energies and that the Froissart bound in the weak sense $|{\cal A}(s,t)|<|s|^2$ is respected at sufficiently large $|s| \rightarrow \infty$ (the fact that at low energy another scaling in $s$ is observed is irrelevant). A non-local UV completion could in principle violate the latter and would evade these considerations, but would in itself be a startling conclusion. \\

We begin in section~\ref{sec:qed} with a review of the standard discussion of positivity bounds as applied to the low energy gravitational Euler-Heisenberg Lagrangian. In section~\ref{scalarQEDbounds} we derive the improved positivity bounds for scalar QED, and in section~\ref{spinorQEDbounds} perform the analogous calculation for spinor QED. Most of the calculational details are saved for the appendices.

\section{Bounds from Euler-Heisenberg}\label{sec:qed}
In the following we consider the theory of QED minimally coupled to gravity, which is itself a low energy EFT. The action for the fermionic (spinor) QED reads
\be\label{qed}
\L_{\rm QED}=\sqrt{-g}\left[\frac{\mpl^2}{2}R-\frac{1}{4}F_{\mu\nu}F^{\mu\nu}- \bar\psi(i\slashed\nabla+m)\psi-eA_\mu\bar\psi\gamma^\mu\psi\right]\,,
\ee
where $\psi$ is the Dirac field, $\bar \psi\equiv\psi^\dagger\gamma^0$, $\slashed\nabla\equiv\gamma^\mu\nabla_\mu$, and $\gamma^\mu=v^{\mu}_a \gamma^a$ are the gamma matrices, $v_{\mu}^a$ the vierbein, and $\nabla$ the covariant derivative with the spin-connection (see appendix \ref{fermionscurved}). We denote by $m$ and $e$ the electron mass and charge respectively.
When  the role of the electron is played by a complex scalar field, the action for scalar QED is then
\be\label{sqed}
\L_{\rm sQED}=\sqrt{-g}\left[\frac{\mpl^2}{2}R-\frac{1}{4}F_{\mu\nu}F^{\mu\nu}-D_\mu\phi D^\mu\phi^\dagger -m^{2}\phi\phi^\dagger\right]\,,
\ee
where $\phi$ is the complex scalar and the gauge-covariant derivative is defined as usual $D_\mu\equiv\partial _\mu-ieA_\mu$.
 Throughout this work we use mostly plus signature $(-,+,+,+)$.

\subsection{Gravitational Euler-Heisenberg effective field theory}
\label{sec:EH_EFT}
Below the electron mass, we may integrate out the heavy electron from \eqref{qed} and \eqref{sqed} respectively. We refer to this as the gravitational Euler-Heisenberg effective field theory. The resulting EFT involves higher derivative interactions between the Maxwell field and graviton and can be parameterized as:
\be\label{IRQED}
\begin{split}
S_{\rm Eul-Heis,1}=\int \d^4x\sqrt{-g}&\left[\frac{\mpl^2}{2}R-\frac{1}{4}F_{\mu\nu}F^{\mu\nu}+\frac{a_1}{m^4}(F_{\mu\nu}F^{\mu\nu})^2+\frac{a_2}{m^4}(F_{\mu\nu}\tilde F^{\mu\nu})^2\right.\\
&\frac{b_1}{m^2}RF_{\mu\nu}F^{\mu\nu}+\frac{b_2}{m^2}R_{\mu\nu}F^{\mu\lambda}F^\nu\,_\lambda+\frac{b_3}{m^2}R_{\mu\nu\lambda\rho}F^{\mu\nu}F^{\lambda\rho}\\
&\left.+c_1R^2+c_2R_{\mu\nu}R^{\mu\nu}+c_3R_{\mu\nu\rho\sigma}R^{\mu\nu\rho\sigma}+\cdots\right]\,,
\end{split}
\ee
where the ellipses designate higher order operators and
where we have defined $\tilde F_{\mu\nu}\equiv\varepsilon_{\mu\nu\alpha\beta}F^{\alpha\beta}/2$.
The form of these operators is the same independently on whether one starts with the spinor or scalar QED, only the exact values of the coefficients $a_i,b_i$ vary. In turn, the $c_i$ couplings appearing in front of the curvature-squared operators are different. These arise even in the case when electron charge $e$ is zero and encode the backreaction of any matter fields on the metric, more precisely the propagator of the spin-2 state. The couplings $c_i$ thus receive contributions from any matter field coupled to gravity and are not solely determined from our QED EFT. The role of these terms is discussed in more detail in section~\ref{higherorder}.\\

The coefficients for the spinor QED are known to be~\cite{Drummond:1979pp,Cheung:2014ega}:
\be\label{coeffs_qed}
\begin{split}
&a_1=\frac{\alpha^2}{90}\,,\qquad a_2=\frac{7\alpha^2}{360}\,,\\
&b_1 = \frac{\alpha}{144\pi}\,,\qquad b_2=-\frac{13\alpha}{360\pi}\,,\qquad b_3 = \frac{\alpha}{360\pi}\,,
\end{split}
\ee
while for scalar QED the coefficients are \cite{Hollowood:2008kq,Shore:2002gw,Cheung:2014ega}
\be\label{coeffs_sqed}
\begin{split}
&a_1=\frac{7\alpha^2}{1440}\,,\qquad a_2=\frac{\alpha^2}{1440}\,,\\
&b_1 = -\frac{\alpha}{288\pi}\,,\qquad b_2=-\frac{\alpha}{360\pi}\,,\qquad b_3 = -\frac{\alpha}{720\pi}\,,
\end{split}
\ee
where $\alpha = e^2/(4\pi)$ is the fine-structure constant. The action \eqref{IRQED}  can be further simplified by expressing the Riemann tensor in terms of the Weyl tensor $C$ and using the lowest order Einstein equations (i.e. performing a field redefinition).  To this order in the EFT, this leads to
\be\label{IRQED2}
\hspace{-0.5cm}\mathcal L_{\rm Eul-Heis,2}=\sqrt{-g}\left[\frac{\mpl^2}{2}R-\frac{1}{4}F_{\mu\nu}F^{\mu\nu}+\frac{a_1'}{m^4}(F_{\mu\nu}F^{\mu\nu})^2+\frac{a'_2}{m^4}(F_{\mu\nu}\tilde F^{\mu\nu})^2+\frac{b_3}{m^2}F_{\mu\nu}F_{\rho\sigma}C^{\mu\nu\rho\sigma}\right]\!,\hspace{-0.3cm}
\ee
where $\tilde F\mn$ is the dual field strength tensor, and (after setting $c_i=0$) the new coefficients are\footnote{Note that these relations differ slightly from those given in Eq. (3.4) of \cite{Cheung:2014ega}. Importantly, there is a sign difference in both $b_2$ and $b_3$ due to the fact that the coefficients $b_i$ change sign under the signature change. The numerical factors coincide if one switches the units, e.g. $1/4\mpl^2=4\pi G/2 = 1/2$, since $4\pi G\equiv 1$ in \cite{Cheung:2014ega}. }
\be\label{aprime}
a_1'=a_1+\frac{1}{4}\frac{m^2}{\mpl^2}b_2+\frac{1}{2}\frac{m^2}{\mpl^2}b_3\,,\qquad a_2'=a_2+\frac{1}{4}\frac{m^2}{\mpl^2}b_2+\frac{1}{2}\frac{m^2}{\mpl^2}b_3\,.
\ee
Notably, both couplings $a_i$ and $b_i$ contribute to the two $F^4$ terms in the action, it is however important to emphasize the difference in their physical origin. For this, let us note that in the gravitational Euler-Heisenberg action \eqref{IRQED2} these arise with different mass scalings in front of the corresponding operators, so that we have
\be
\frac{a_i}{m^4}\sim\frac{e^4}{m^4}\,,\qquad\frac{m^2}{\mpl^2}\frac{b_i}{m^4}\sim\frac{e^2}{m^2\mpl^2}\,.
\ee
The appearance of the inverse powers of $\mpl$ in the $b$-terms indicate that the scattering processes leading to the low energy $F^4$ interactions are different in the two cases. The couplings $a_i$ are generated by four-photon scatterings involving only electron exchange, (shown on the first line of Fig.~\ref{fig:scattering2} or first diagram of Fig.~\ref{fig:scattering3}). The couplings $b_i$ in turn are generated by gravitational four-photon scattering involving a massless graviton exchange as shown on the second line of Fig.~\ref{fig:scattering2} (or last three diagrams of Fig.~\ref{fig:scattering3}).

\begin{figure}[t]
    \centering
\includegraphics[width=10cm]{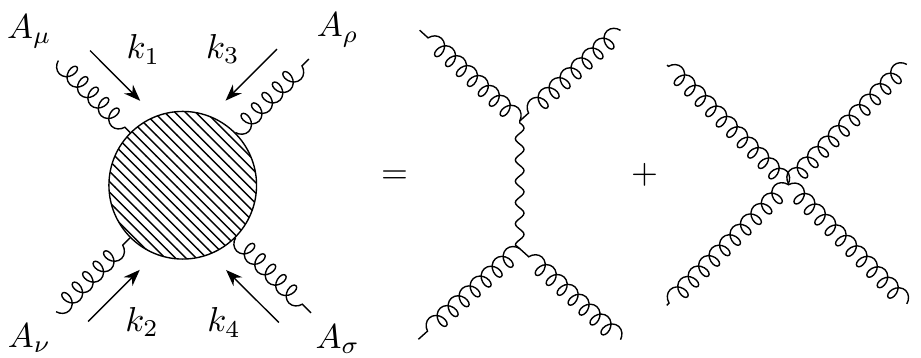}
\caption{The $AA\to AA$ $t$-channel scattering in the gravitational Euler-Heisenberg theory. The wiggly line stands for the vector field $A_\mu$. The exchanged wavy line stands for the graviton $h_{\mu\nu}$. }
\label{fig:scattering}
\end{figure}

\subsection{Positivity Bounds from the Euler-Heisenberg EFT}\label{IRpositivity}
The leading contribution to the four photon $AA\to AA$ scattering amplitude in the gravitational Euler-Heisenberg theory below the electron mass \eqref{IRQED2} comes from the scattering processes shown in Fig.~\ref{fig:scattering}. Although not explicit in the diagrams, $b_3$ enters through a modified graviton-photon-photon vertex. Consistently with the previous literature, we find the following results for the various helicity configurations of the ingoing and outgoing particles (written in an all ingoing convention):
\be\label{ampIR}
\begin{split}
&\mathcal  A_{\rm Eul-Heis}(++++)=\mathcal  A_{\rm Eul-Heis}(----)=\frac{8(a'_1-a'_2)}{m^4}\left(s^2+t^2+u^2\right)\,,\\
&\mathcal  A_{\rm Eul-Heis}(++--)=\mathcal  A_{\rm Eul-Heis}(--++)=\frac{s^4}{\mpl^2stu}+\frac{8(a'_1+a'_2)}{m^4}s^2\,,\\
&\mathcal  A_{\rm Eul-Heis}(+-+-)=\mathcal  A_{\rm Eul-Heis}(-+-+)=\frac{t^4}{\mpl^2stu}+\frac{8(a'_1+a'_2)}{m^4}t^2\,,\\
&\mathcal  A_{\rm Eul-Heis}(+--+)=\mathcal  A_{\rm Eul-Heis}(-++-)=\frac{u^4}{\mpl^2stu}+\frac{8(a'_1+a'_2)}{m^4}u^2\,.
\end{split}
\ee
The $b_3$ interaction vertex only contributes to the $\mathcal  A_{\rm Eul-Heis}(+++-)$, $\mathcal  A_{\rm Eul-Heis}(---+)$ etc. amplitudes  as \cite{Cheung:2014ega}:
\be\label{ampb3}
\mathcal  A_{\rm Eul-Heis}(+++-)=\mathcal  A_{\rm Eul-Heis}(---+)=\frac{b_3}{\mpl^2m^2}(s^2+t^2+u^2)\,.
\ee
These amplitudes respect $s$-$u$ crossing symmetry in the sense
\be
\mathcal  A(\lambda_1 \lambda_2 \lambda_3 \lambda_4)(s,t,u) = \mathcal  A(\lambda_1 \lambda_4 \lambda_3 \lambda_2)(u,t,s) \, .
\ee
As expected, the amplitudes \eqref{ampIR} involve the infamous $t$-channel pole diverging in the forward limit thus formally invalidating any analyticity arguments that would lead to the positivity bounds. Interestingly, in \cite{Cheung:2014ega} it was proven that upon discarding the massless graviton pole and after symmetrizing the scattering amplitudes above, the positivity bounds imply
\be\label{boundIR0}
a'_1+a'_2>0\,.
\ee
Alternatively, this result may also be obtained by  analyzing the elastic amplitude $\mathcal  A(++--)\equiv \mathcal  A(++ \rightarrow ++)$ alone. Inserting the expressions of the coefficients \eqref{aprime} we get
\be\label{boundIR}
a_1+a_2+\frac{m^2}{\mpl^2}\left(\frac{b_2}{2}+b_3\right)>0\,.
\ee
As discussed earlier, the exact values of the coefficients $a_i$ and $b_i$ are known from the QED EFT \eqref{IRQED2} and are given in Eqs.~\eqref{coeffs_qed} and \eqref{coeffs_sqed}. For the scalar QED this implies\footnote{The relations \eref{wgc1} and \eqref{wgc2} are given for $c_i=0$ whereas \cite{Cheung:2014ega} also accounts for the non-zero $c_i$. The implications of non-zero $c_i$, which contribute at order $1/\mpl^4$ in the amplitudes, are discussed in section~\ref{higherorder}.}
\be\label{wgc1}
\frac{e^4}{2880\mpl^2\pi^2}\left(-2\frac{m^2}{e^2}+\mpl^2\right)>0\,,
\ee
while for the spinor QED this leads to
\be\label{wgc2}
\frac{e^4}{5760\mpl^2\pi^2}\left(-24\frac{m^2}{e^2}+11\mpl^2\right)>0\,.
\ee
Taking these bounds at their face value one would be tempted to conclude that these imply the weak-gravity type of bounds on the charge-to-mass ratio, i.e. that $e/m\gtrsim\sqrt{2}/\mpl$, which was one of the remarkable points presented in \cite{Cheung:2014ega}. However as we shall see below, the previous bounds rely on known positive QED contributions, namely that from the non-gravitational electron loop. However the raison d'\^etre of positivity bounds is to probe the unknown UV contributions. Any known  contributions from the EFT can and should be removed by means of the improved positivity bounds, as we describe below, before any physical conclusions are derived.\\

The bounds \eqref{boundIR0} are not the only bounds that may be derived assuming the $t$-channel pole may be discarded, we may also consider states of indefinite polarization which mix in information about $b_3$. For instance, taking the incoming polarization state to be $|+ \rangle \otimes \frac{1}{\sqrt{2}} ( | + \rangle \pm |- \rangle)$ the positivity bound becomes $4 (a_1' +a_2')> m^2|b_3|/\mpl^2 $. For specific indefinite polarization states corresponding to those that are natural from compactification to 3D, we may then recover for example the bounds argued for in \cite{Bellazzini:2019xts}. In our current notation these are the statements that
\be
4 a_1' > \frac{m^2}{\mpl^2}|b_3|  \, , \quad a_2' >0 \, ,
\ee
which are stronger and therefore include \eqref{boundIR}. Once again, taken at face value for QED minimally coupled to gravity, we would be led to a similar conclusion about the charge-to-mass ratio in order to satisfy them, a conclusion that would be premature.\\

Before proceeding we note that the bound discussed in  \cite{Cheung:2014ega} has been countered in the case of 3D by the discussion of \cite{Chen:2019qvr} which make use of extended positivity bounds of \cite{AHH}, leading to an opposing bound on the charge-to-mass ratio. This parallels some of the discussion in what follows for 4D, although we shall make use of the improved positivity bounds which allows us to infer a bound on the cutoff of the EFT and avoid the need to focus on the high powers of $s$ in the expansion of the amplitude.

\section{Bounds from scalar QED coupled to gravity} \label{scalarQEDbounds}

Our goal is to extend the argumentation of the previous section, whereby, instead of applying the positivity bounds to the
 Euler-Heisenberg Lagrangian, we shall apply them
directly to QED minimally coupled to gravity - itself treated as a low energy EFT. The new feature is that the resulting EFT is valid at and above the mass of the electron (up to the EFT cutoff $\Lambda_c$), and so we may use the `knowledge' of electron loop contributions to `improve' the positivity bounds. Before we do this we outline in more detail the improved positivity bounds in the next subsection.

\begin{figure}[t]
    \centering
\includegraphics[width=13cm]{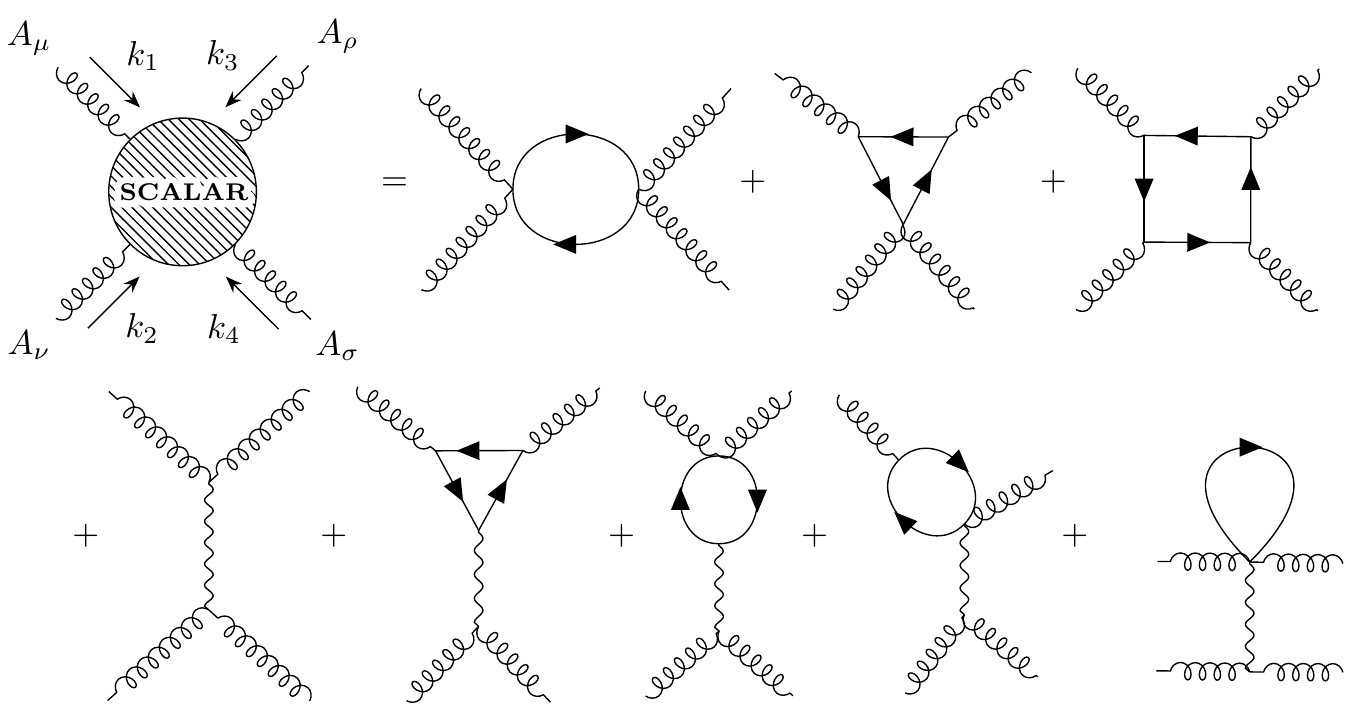}
\caption{The $AA\to AA$ scattering in scalar QED due to non-gravitational interactions (first line) and gravitational interactions to order $1/\mpl^2$ (second line). The wiggly line stands for the vector field $A_\mu$ and the solid line stands for the scalar field $\phi$. The arrows depict the direction of the charge flow. We do not show all the crossed versions of the diagrams. }
\label{fig:scattering2}
\end{figure}

\subsection{Improved positivity and dispersion relations}

The fixed $t$ dispersion relation for the pole-subtracted amplitude $\tilde{\mathcal{A}}(s,t,u)$ can be written in a maximally $s$-$u$ crossing symmetric way as
\be
\begin{split}
\tilde{\mathcal{A}}(s,t,u)=a_1(t)+a_2(t)s+\frac{s^2}{\pi}\int_{0}^{\infty}\rd s'\,\frac{\textrm{Disc}_s\mathcal{A}(s',t,u')}{s^{'2}(s'-s)}+\frac{u^2}{\pi}\int_{0}^{\infty}\rd u'\,\frac{\textrm{Disc}_{u}\mathcal{A}(s',t,u')}{u^{'2}(u'-u)}\,,
\end{split}
\ee
where $s'+u'+t=0$ for massless photons, and the discontinuities here are with respect to the Mandelstam variable that corresponds to the center of mass (CoM) energy squared of either the $s$ or $u$ channel
\be
2\ri\textrm{Disc}_s\mathcal{A}(s,t,u)\equiv\mathcal{A}(s+\ri\epsilon,t,u=-s-t-\ri\epsilon)-\mathcal{A}(s-\ri\epsilon,t,u=-s-t+\ri\epsilon)\,,
\ee
and
\be
2\ri\textrm{Disc}_u\mathcal{A}(s,t,u)\equiv\mathcal{A}(s=-u-t-\ri\epsilon,t,u+\ri\epsilon)-\mathcal{A}(s=-u-t+\ri\epsilon,t,u-\ri\epsilon)\,.
\ee
Explicitly, crossing symmetry implies that
\be
\mathcal{A}_{\lambda_1 \lambda_2 \rightarrow \lambda_3 \lambda_4}(s,0,u) = \mathcal{A}_{\lambda_1 -\lambda_4 \rightarrow \lambda_3 -\lambda_2}(u,0,s)\,,
\ee
and so the left-hand $u$-channel discontinuity is defined so that
\be
\textrm{Disc}_u \mathcal{A}_{\lambda_1 \lambda_2 \rightarrow \lambda_3 \lambda_4}(s,0,u) = \textrm{Disc}_u \mathcal{A}_{\lambda_1 -\lambda_4 \rightarrow \lambda_3 -\lambda_2}(u,0,s) = \big[ \textrm{Disc}_s \mathcal{A}_{\lambda_1 -\lambda_4 \rightarrow \lambda_3 -\lambda_2}(s,0,u) \big]_{u \leftrightarrow s} \, ,
\ee
which is just the standard right-hand discontinuity of the crossed process $A + \bar D \rightarrow C + \bar B$ (associated with $A+B \rightarrow C+D$). The physical discontinuities are therefore positive in the forward limit for elastic scattering by unitarity on both the right hand and left hand cuts, leading to the forward limit positivity bound,
\be\label{forwardlimitbound}
\begin{split}
\partial_s^2\tilde{\mathcal{A}}(0,0,0)=\frac{2}{\pi}\int_{0}^{\infty}\rd s'\,\frac{\textrm{Disc}_s\mathcal{A}(s',0,u')}{s^{'3}}+\frac{2}{\pi}\int_{0}^{\infty}\rd u'\,\frac{\textrm{Disc}_{u}\mathcal{A}(s',0,u')}{u^{'3}}>0\,.
\end{split}
\ee
This positivity bound can be improved by then subtracting a known positive contribution to the discontinuities from both sides of the dispersion relation. This discontinuity can be computed in the EFT (e.g. QED in our case), giving a result that can be trusted well below its cutoff scale $\Lambda_c$, hence the integrals over $s'$ and $u'$ must be cutoff at $\epsilon^2\Lambda_c^2$ with $\epsilon\ll1$. This can then be achieved by a split
\be
\begin{split}
\textrm{Disc}_{s}\mathcal{A}(s',0,u')&=\textrm{Disc}_{s}\mathcal{A}(s',0,u')\theta(\epsilon^2\Lambda_c^2-s)+\textrm{Disc}_{s}\mathcal{A}(s',0,u')\theta(s-\epsilon^2\Lambda_c^2)\\
\textrm{Disc}_{u}\mathcal{A}(s',0,u')&=\textrm{Disc}_{u}\mathcal{A}(s',0,u')\theta(\epsilon^2\Lambda_c^2-u)+\textrm{Disc}_{u}\mathcal{A}(s',0,u')\theta(u-\epsilon^2\Lambda_c^2)\,,
\end{split}
\ee
where the first term on the RHS is regarded as the `known' part of the discontinuity, and both known and unknown pieces are positive separately.
We may then define the improved scattering amplitude ${\mathcal A}^{\rm imp}(s,t,u)$ via \cite{deRham:2017xox,deRham:2017imi}.
\be\label{improvedamplitude}
{{\mathcal{A}}^{\rm imp}}(s,t,u)\equiv \tilde{\mathcal{A}}(s,t,u) - \frac{s^2}{\pi}\int_{0}^{\epsilon^2\Lambda_c^2}\rd s'\,\frac{\textrm{Disc}_s\mathcal{A}(s',t,u')}{s^{'2}(s'-s)}-\frac{u^2}{\pi}\int_{0}^{\epsilon^2\Lambda_c^2}\rd u'\,\frac{\textrm{Disc}_{u}\mathcal{A}(s',t,u')}{u^{'2}(u'-u)} \, .
\ee
Crucially ${\mathcal A}^{\rm imp}(s,t,u)$ has the same analytic structure as $\tilde{\mathcal{A}}(s,t,u) $ except that by construction the branch cuts now start at $s'=\epsilon^2\Lambda_c^2$ and $u'=\epsilon^2\Lambda_c^2$ . We may then derive improved positivity bounds from $ \tilde{\mathcal A}^{\rm imp}(s,t,u) $ in the same manner in which they are derived from ${\mathcal A}^{\rm imp}(s,t,u)$, in particular  leading to the forward limit bound
\be
\begin{split}
\partial_s^2{{\mathcal{A}}^{\rm imp}}(0,0,0)=\partial_s^2\tilde{\mathcal{A}}(0,0,0)-\frac{2}{\pi}\int_{0}^{\epsilon^2\Lambda_c^2}\rd s'\,\frac{\textrm{Disc}_{s}\mathcal{A}(s',0,u')}{s^{'3}}-\frac{2}{\pi}\int_{0}^{\epsilon^2\Lambda_c^2}\rd u'\,\frac{\textrm{Disc}_{u}\mathcal{A}(s',0,u')}{u^{'3}}>0 \, .
\end{split}
\ee
To proceed we need to know not only the low energy expansion of the amplitude, but also the low energy discontinuities. These receive contributions from both non-gravitational diagrams and gravitational ones, and we shall deal with each of these in turn.

\subsection{Discontinuities of non-gravitational diagrams}

The full set of diagrams for scalar QED that contribute to the one-loop 4-photon amplitude to order $1/\mpl^2$, including graviton exchange, are given in Fig.~\ref{fig:scattering2}. In general the discontinuities can be inferred by unitarity cuts, however we choose to derive them directly from the amplitudes provided in appendix~\ref{1loopscalarqed}. The discontinuities are calculated within the domain relevant to the dispersion relation, namely the physical region which for e.g. $\disc_s\cA(s',0,u')$ is  $s'\geq0$. In general it is necessary to keep track of both the discontinuities of the original process $A+B \rightarrow C+D$ and the crossed process $A+ \bar D \rightarrow C+ \bar B$. We use  the results and notation of appendix \ref{scalarQEDnongrav}. Focusing for now on the non-gravitational contributions (i.e. those with no internal graviton lines), we denote by  $\mathcal{A}_n$ the contributions to the amplitude arising from diagrams with $n$ internal $\phi$ propagators. For scalar QED the relevant discontinuities from individual Feynman diagrams are respectively:\\

\noindent $\bullet$ For two internal lines,
\be
\begin{split}
\disc_s\cA_2(s,0,u)&=\frac{e^4}{4\pi}\vep_{12}\vep_{34}\sqrt{\frac{s-4m^2}{s}}\theta(s-4m^2) \, , \\
\end{split}\nn
\ee
$\bullet$ For three internal lines,
\be
\begin{split}
\disc_s\cA_3(s,0,u)&=-\frac{e^4}{\pi}\vep_{12}\vep_{34}\left(\frac{\sqrt{s(s-4m^2)}+2m^2\ln\left(\frac{2m^2}{s-2m^2+\sqrt{s(s-4m^2)}}\right)}{2s}\right)\theta(s-4m^2)\, ,\\
\end{split}
\ee
$\bullet$ And finally for four internal lines,
\be
\begin{split}
    \disc_s\cA_4(s,0,u)&=\frac{e^4 }{8\pi s^2}\Bigg\{\sqrt{s(s-4m^2)}(s+2m^2)-\frac{4s}{m^2}\left(\ln{4}+3\ln{\frac{s}{m^2}}-4\ln{\frac{s-\sqrt{s(s-4m^2)}}{m^2}}\right)   \\
    &-2(m^2+4s)\ln{\left(1-\sqrt{\frac{s-4m^2}{s}}\right)}+2\ln{\left(-3-\sqrt{\frac{s-4m^2}{s}}+\frac{s}{m^2}+\frac{\sqrt{s(s-4m^2)}}{m^2}\right)}\Bigg\}\\&\times\theta(s-4m^2)\times\left(\vep_{12}\vep_{34}+\vep_{14}\vep_{23}+\vep_{13}\vep_{24}\right) \, .
\end{split}\nn
\ee
In all cases the associated $u$-channel discontinuities can be inferred from
\be
\disc_u\cA_n(s,0,u) = [\disc_s\cA_n(s,0,u)]_{s \leftrightarrow u , 2 \leftrightarrow 4} \, .
\ee
The total amplitude is the sum of all contributions $\cA(s,t,u)=Z_A^2 \cA_{\rm tree}(s,t,u)+\cA_2(s,t,u)+\cA_3(s,t,u)+\cA_4(s,t,u)$, accounting for wavefunction renormalization, and so the discontinuities combine accordingly.

\subsection*{Checking against the dispersion relation}
As a simple consistency check, we can verify the expression \eqref{forwardlimitbound} for the second derivative of the dispersion relation. Direct integration of the discontinuities gives
\be
\begin{split}
&\frac{2}{\pi}\int_{0}^{\infty}\rd s'\,\frac{\textrm{Disc}_s\mathcal{A}(s',0,u')}{s^{'3}}+\frac{2}{\pi}\int_{0}^{\infty}\rd u'\,\frac{\textrm{Disc}_{u}\mathcal{A}(s',0,u')}{u^{'3}}\\
&=\frac{e^4}{240\pi^2 m^4}\left(\vep_{12}\vep_{34}+\vep_{14}\vep_{23}+\frac13\vep_{13}\vep_{24}\right)\,,
\end{split}
\ee
whereas taking the derivative directly of the non-gravitational amplitude gives,
\be
\partial_s^2 \tilde{\mathcal{A}}(0,0,0)=\frac{e^4}{240\pi^2 m^4}\left(\vep_{12}\vep_{34}+\vep_{14}\vep_{23}+\frac13\vep_{13}\vep_{24}\right)\,,
\ee
confirming the validity of the dispersion relation with two subtractions in the absence of gravity.

\subsection{Discontinuities of gravitational diagrams}
The gravitational diagrams for scalar QED that contribute to order $1/\mpl^2$ are computed in appendix~\ref{ScalarQEDoneloop} and contain individual Feynman diagram contributions labelled $\fa,\fb,\fc,\fd$ in Fig.~\ref{fig:scattering4}. We find that the type $\fc,\fd$ diagrams do not produce any discontinuity. This is because the denominator of the loop integrand has strictly positive real part. We find the $\fb$ type diagrams also has zero imaginary part, so we can focus solely on the type-$\fa$ diagrams.
We shall define the following scattering configurations
\ba
\begin{array}{lccc}
 \text{Configuration \RNum{1}}: & \quad  ++--\, & \, \equiv \, & \, ++ \rightarrow ++  \\
 \text{Configuration \RNum{2}}: & \quad +--+ \, &\,  \equiv \, & \, +- \rightarrow +-  \\
 \text{Configuration \RNum{3}}: & \quad +-+- \, &\,  \equiv \,  & \,  +- \rightarrow -+
\end{array}\nn
\ea
and focus only on these for illustrative purposes. The first two configurations are elastic so positivity bounds apply to them.

\paragraph{Configuration \RNum{1}}
The loop diagrams with one graviton exchange have the following discontinuities.
\be
\disc_s\cA_{\text{\RNum{1}}}(s,0,u)=0\,,
\ee
\be
\disc_u\cA_{\text{\RNum{1}}}(s,0,u)=\frac{e^2}{24\pi\mpl^2 u}\left((10 m^2-u) \sqrt{u \left(u-4 m^2\right)}-24 m^4 \tanh^{-1}\left(\sqrt{\frac{u-4 m^2}{u}}\right)\right)\theta(u-4m^2)\,.
\ee
Note that this discontinuity is strictly negative in the physical region. This does not contradict unitarity since this is a perturbative gravitational correction to an already positive non-gravitational discontinuity.
\paragraph{Configuration \RNum{2}}

\be
\disc_s\cA_{\text{\RNum{2}}}(s,0,u)=\frac{e^2}{24\pi \mpl^2 s}\left((10 m^2-s) \sqrt{s \left(s-4 m^2\right)}-24 m^4 \tanh^{-1}\left(\sqrt{\frac{s-4 m^2}{s}}\right)\right)\theta(s-4m^2)\,.
\ee
Note that this discontinuity is also strictly negative, while the $u$-channel contribution cancels,
\be
\disc_u\cA_{\text{\RNum{2}}}(s,0,u)=0\,.
\ee

\paragraph{Configuration \RNum{3}}
The forward limit of this helicity configuration has zero discontinuity which agrees with the gravitational Euler-Heisenberg result as the gravitational part of the amplitude in this configuration is zero in the forward limit.

\subsection*{Checking against the dispersion relation}
If it were the case that QED coupled to gravity still respected the Jin-Martin version of the Froissart bound to one-loop level, i.e. $|A(s,t)|<|s|^2$, then it would  still be possible to write a dispersion relation for the scattering amplitude with two subtractions. Furthermore if this were the case it would be possible to use the improved positivity bound to remove even the gravitational contributions. Fortunately this is not the case and it is this very fact that will lead to our central result. For scalar QED in configuration \RNum{1} the gravitational contribution to the amplitude gives,
\be
\cA_{\textrm{\RNum{1}}}''(0)=-\frac{e^2}{90\pi^2 m^2\mpl^2}
\ee
whereas the usual dispersion integrals give,
\be
\frac{2}{\pi}\int_{0}^{\infty}\rd s'\,\frac{\disc_s\cA_{\text{\RNum{1}}}(s',0,u')}{s^{'3}}+(s\xleftrightarrow[]{} u)=-\frac{e^2 }{180m^2\pi^2\mpl^2}\,.
\ee
By contrast higher derivatives of the dispersion relation do match which is to be expected since we {\it can} write a dispersion relation for the one-loop gravitational contribution with 3 subtractions, and for any $n\ge 3$, so we do expect the following to hold:
\be
\p_s^n\mathcal{A}(0,0,0)=\frac{n!}{\pi}\int_{0}^{\infty}\rd s'\,\frac{\disc_s\cA(s',0,u')}{s^{'\,n+1}}+\frac{(-1)^n n!}{\pi}\int\rd u'\,\frac{\disc_u\cA(s',0,u')}{u^{'\,n+1}}\,.
\ee
Indeed for configuration I the integrals on the RHS give,
\be
\label{eq:discint}
0+\frac{(-1)^n n!}{\pi}\int\rd u'\,\frac{\disc_u\cA_{\textrm{\RNum{1}}}(s',0,u')}{u^{'\,n+1}}=\frac{e^2 (-1)^n 2^{-2 (n+1)} n! \csc (\pi  n)}{\sqrt{\pi } \mpl^2 m^{2(n-1)}(n+1) \Gamma (2-n) \Gamma \left(n+\frac{3}{2}\right)}\,,
\ee
whereas from taking derivatives of the amplitude for the LHS we obtain
\be
\cA_{\textrm{\RNum{1}}}^{(n)}(0,0,0)=\frac{e^2 (-1)^{n+1} \Gamma (n-1) \Gamma (n+1) \Gamma (n+2)}{\mpl^2 m^{2(n-1)}\pi ^2 (n+1) \Gamma (2 n+3)}\,,
\ee
which is in fact equal to the dispersion integral result, so the dispersion relation holds for all $n\geq 3$.\\

It is worth stressing that the fact that the low energy amplitude computed within the EFT does not satisfy the
Jin-Martin version of the Froissart bound at the one-loop level
 does not in any way imply that the full UV amplitude would itself need to violate Froissart. It is indeed inevitable in an EFT that amplitudes computed to finite order in an energy expansion grow `too fast', indicating only the breakdown of the effective theory. It is precisely because of this fact that positivity bounds are so powerful.

\subsection{Improved positivity bounds}

We now have all the essential ingredients needed to derive our main result. Assuming that the improved positivity bounds can be applied with the $t$-channel pole discarded, then the fact that the non-gravitational amplitudes for QED do respect Froissart at one-loop, but the gravitational corrections (as computed within the QED EFT) do not, will enforce a non-trivial bound on the cutoff of the low energy effective theory. Let us focus on the configuration I amplitude which is elastic in polarizations $++ \rightarrow ++$.

The $s$-channel integral is,
\be
\begin{split}
    &\frac{2}{\pi}\int_{4m^2}^{\epsilon^2\Lambda_c^2}\rd s\,\frac{\disc_s\cA(s,0,u)}{s^3}=\\ &\frac{2e^2}{\pi^2}\int_{4m^2}^{\epsilon^2\Lambda_c^2}\frac{\rd s}{s^3}\,\Bigg[\frac{e^2}{4}\sqrt{\frac{s-4m^2}{s}}-e^2\left(\frac{\sqrt{s(s-4m^2)}+2m^2\ln\left(\frac{2m^2}{s-2m^2+\sqrt{s(s-4m^2)}}\right)}{2s}\right)\\
    &\qquad\quad+\frac{e^2 }{4 s^2}\Bigg\{\sqrt{s(s-4m^2)}(s+2m^2)-\frac{4s}{m^2}\left(\ln{4}+3\ln{\frac{s}{m^2}}-4\ln{\frac{s-\sqrt{s(s-4m^2)}}{m^2}}\right)\\
    &-2(m^2+4s)\ln{\left(1-\sqrt{\frac{s-4m^2}{s}}\right)}+2\ln{\left(-3-\sqrt{\frac{s-4m^2}{s}}+\frac{s}{m^2}+\frac{\sqrt{s(s-4m^2)}}{m^2}\right)}\Bigg\} \Bigg]\,,
\end{split}
\ee
while the $u$-channel integral is,
\be
\begin{split}
    &\frac{2}{\pi}\int_{4m^2}^{\epsilon^2\Lambda_c^2}\rd u\,\frac{\disc_u\cA(s,0,u)}{u^3}=\\ &\frac{2e^2}{\pi^2}\int_{4m^2}^{\epsilon^2\Lambda_c^2}\frac{\rd u}{u^3}\,\Bigg[\frac{e^2 }{4 u^2}\Bigg\{\sqrt{u(u-4m^2)}(u+2m^2)-\frac{4u}{m^2}\left(\ln{4}+3\ln{\frac{u}{m^2}}-4\ln{\frac{u-\sqrt{u(u-4m^2)}}{m^2}}\right)\hspace{-1cm}\\
    &-2(m^2+4u)\ln{\left(1-\sqrt{\frac{u-4m^2}{u}}\right)}+2\ln{\left(-3-\sqrt{\frac{u-4m^2}{u}}+\frac{u}{m^2}+\frac{\sqrt{u(u-4m^2)}}{m^2}\right)}\Bigg\}\\
    &\qquad\quad+\frac{1}{24\mpl^2 u}\left((10 m^2-u) \sqrt{u \left(u-4 m^2\right)}-24 m^4 \tanh^{-1}\left(\sqrt{\frac{u-4 m^2}{u}}\right)\right)\Bigg]\,.
\end{split}
\ee
Inserting the discontinuities into the improved positivity bounds and expanding the integrals given the necessary assumption $m \ll \epsilon \Lambda_c$ leads to
\be
\begin{split}
0<&\,\partial_s^2\tilde{\mathcal{A}}(0,0,0)-\frac{2}{\pi}\int_{0}^{\epsilon^2\Lambda_c^2}\rd s'\,\frac{\textrm{Disc}_{s}\mathcal{A}(s',0,u')}{s^{'3}}-\frac{2}{\pi}\int_{0}^{\epsilon^2\Lambda_c^2}\rd u'\,\frac{\textrm{Disc}_{u}\mathcal{A}(s',0,u')}{u^{'3}} \, ,\\
0<&\,\frac{e^4}{4 \pi ^2 \Lambda_c ^4 \epsilon ^4}+\frac{e^2 m^2}{2 \pi ^2  \mpl^2 \Lambda_c ^4 \epsilon ^4}-\frac{e^2}{180 \pi ^2 m^2 \mpl^2}-\frac{e^2}{12 \pi ^2  \mpl^2 \Lambda_c ^2 \epsilon ^2}+ \dots\, .
\end{split}
\ee
Given the assumed EFT hierarchy $m \ll \epsilon \Lambda_c \ll \mpl$ it is sufficient to approximate this as
\be\label{bound101}
\,\frac{e^4}{4 \pi ^2 \epsilon ^4\Lambda_c ^4 } -\frac{e^2}{180 \pi ^2 m^2 \mpl^2} >0\, .
\ee
The first term is the contribution from non-gravitational diagrams which in the absence of gravity can be removed by sending $\Lambda_c \rightarrow \infty$, reflecting the statement that QED in flat space automatically satisfies positivity bounds to one-loop. The second term is the distinctively negative gravitational contribution which arises from the non-Froissart growth of the one-loop amplitudes\footnote{We stress again that this does not imply that the UV amplitudes violate the weak Froissart bound $|{\cal A}(s,t)|<|s|^2$.}.
This positivity bound may be most cleanly interpreted as a bound on the cutoff of the effective theory:
\be
\epsilon\Lambda_c\lesssim  (\,e m \mpl )^{1/2}\,,
\ee
which is to say that if we take the positivity bound with $t$-channel pole discarded seriously, QED cannot be minimally coupled to gravity without introducing new physics at or below the scale $\Lambda_{\rm new} \sim  (\,e m \mpl )^{1/2}$. This is significantly lower than the scale $e \mpl$ implied by the weak gravity conjecture \cite{ArkaniHamed:2006dz}, and indeed by the Euler-Heisenberg bounds derived in section \ref{IRpositivity}. The strength of this result is due to the fact that we can remove the known QED contributions to the positivity bounds from the electron loops, up to the cutoff scale $\epsilon \Lambda_c$, giving us a much more constraining condition. This result exactly parallels similar conclusions derived for toy scalar field theories coupled to gravity in a previous work \cite{Alberte:2020jsk}. The present result is however cleaner since
(a) we do not rely on spectator fields, (b) the form of the QED lagrangian is more strongly constrained by gauge invariance and (c) we make no assumption on the types of operators that would arise at the cutoff. When the inequality \eqref{bound101} comes close to being saturated we should also worry about higher order corrections in $1/\mpl^2$. These will be considered in section~\ref{higherorder}.

 \section{Bounds from spinor QED coupled to gravity} \label{spinorQEDbounds}

\begin{figure}[t]
    \centering
\includegraphics[width=13cm]{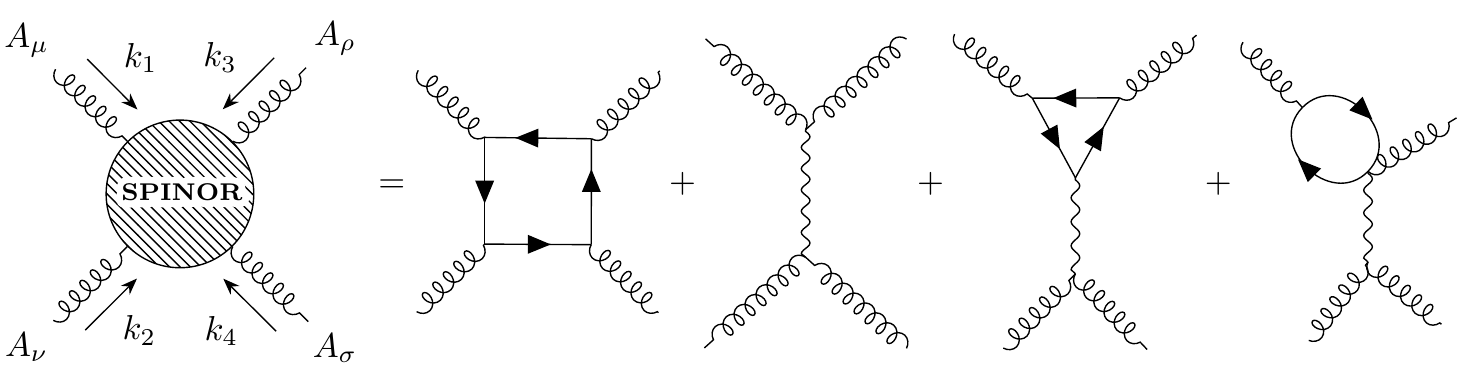}
\caption{The $AA\to AA$ scattering in spinor QED due to non-gravitational interactions (first line) and gravitational interactions to order $1/\mpl^2$ (second line). The wiggly line stands for the vector field $A_\mu$ and the solid line stands for the fermion $\psi$. The arrows depict the direction of the charge flow. We do not show all the crossed versions of the diagrams. }
\label{fig:scattering3}
\end{figure}

The discussion for spinor QED closely parallels that for the scalar QED with the only difference being numerical factors. We sketch the essential arguments leaving the amplitude calculation details to appendix \ref{1loopspinorqed}. The number of diagrams contributing to the 4-photon amplitude at one-loop level and to order $1/\mpl^2$ is significantly fewer as seen in Fig.~\ref{fig:scattering3}.

\subsection{Discontinuities of non-gravitational diagrams}

As shown in Fig.~\ref{fig:scattering3} the only non-gravitational diagram is the `box' diagram. The relevant amplitude discontinuities are given in appendix~\ref{spinornongrav} in the $s$ and $u$ channel. For the first two  polarization configurations they are respectively
\be\nn
\begin{split}
\disc_{s}\cA^{\text{\RNum{1}}}_{\textrm{box}}&=\frac{e^4\pi  \theta \left(s-4 m^2\right)\left(s-2 m^2\right)}{2\pi^2 s^2} \left( \sqrt{s \left(s-4 m^2\right)}-2 m^2 \log \left(\frac{s-\sqrt{s \left(s-4 m^2\right)}}{s+ \sqrt{s \left(s-4 m^2\right)}}\right)\right)\\
\disc_{u}\cA^{\text{\RNum{1}}}_{\textrm{box}}&=\frac{e^4\pi\theta \left(u-4 m^2\right)}{2\pi^2 u^2} \left(2 \left(-m^2-u\right) \sqrt{u \left(u-4 m^2\right)}+\left(4 m^4-2 m^2 u-u^2\right) \log \left(\frac{u-\sqrt{u \left(u-4 m^2\right)}}{u+\sqrt{u \left(u-4 m^2\right)}}\right)\right)\,,
\end{split}
\ee
and
\be\nn
\begin{split}
\disc_{s}\cA^{\text{\RNum{2}}}_{\textrm{box}}&=\frac{e^4\pi\theta \left(s-4 m^2\right)}{2 \pi^2  s^2} \left(2 \left(-m^2-s\right) \sqrt{s \left(s-4 m^2\right)}+\left(4 m^4-2 m^2 s-s^2\right) \log \left(\frac{s-\sqrt{s \left(s-4 m^2\right)}}{s+\sqrt{s \left(s-4 m^2\right)}}\right)\right)\,,\\
\disc_{u}\cA^{\text{\RNum{2}}}_{\textrm{box}}&=\frac{e^4\pi \theta \left(u-4 m^2\right)\left(u-2m^2\right)}{2 \pi^2  u^2} \left(\sqrt{u \left(u-4 m^2\right)}-2 m^2 \log \left(\frac{u-\sqrt{u \left(u-4 m^2\right)}}{u+\sqrt{u \left(u-4 m^2\right)}}\right)\right)\,.
\end{split}
\ee
All the above discontinuities are positive as required by unitarity. We may confirm the validity of the dispersion relation with two subtractions by demonstrating that
\be
\begin{split}
\partial_s^2\tilde{\mathcal{A}}(0,0,0)=\frac{2}{\pi}\int_{0}^{\infty}\rd s'\,\frac{\textrm{Disc}_s\mathcal{A}(s',0,u')}{s^{'3}}+\frac{2}{\pi}\int_{0}^{\infty}\rd u'\,\frac{\textrm{Disc}_{u}\mathcal{A}(s',0,u')}{u^{'3}}=\frac{11 e^4}{360\pi^2 m^4}\,,
\end{split}
\ee
as required,  confirming the discontinuities above for both chosen helicity configurations.

 \subsection{Discontinuities of gravitational diagrams}

For spinor QED the only gravitational discontinuities come from the type $\fa$ diagrams. The discontinuities of these diagrams are negative and are given by,
\be
\begin{split}\nn
\disc_{u}\cA^{\textrm{\RNum{1}}}&=-\frac{e^2}{6 \pi \mpl^2 u} \theta \left(u-4 m^2\right) \left(\left(5 m^2+u\right) \sqrt{u \left(u-4 m^2\right)}+3 m^2 \left(2 m^2+u\right) \log \left(\frac{u-\sqrt{u \left(u-4 m^2\right)}}{u+\sqrt{u \left(u-4 m^2\right)}}\right)\right)\,,\\
\disc_{s}\cA^{\textrm{\RNum{1}}}&=0\,,\\
\disc_{s}\cA^{\textrm{\RNum{2}}}&=-\frac{e^2}{6 \pi \mpl^2 s} \theta \left(s-4 m^2\right) \left(\left(5 m^2+s\right) \sqrt{s \left(s-4 m^2\right)}+3 m^2 \left(2 m^2+s\right) \log \left(\frac{s-\sqrt{s \left(s-4 m^2\right)}}{s+\sqrt{s \left(s-4 m^2\right)}}\right)\right)\,,\\
\disc_{u}\cA^{\textrm{\RNum{2}}}&=0\,.
\end{split}
\ee
As before the negativity of these discontinuities is not in contradiction with unitarity since these are $\mpl^2$ suppressed corrections to the positive non-gravitational discontinuities. Here again, one can explicitly check that this discontinuity is consistent with the relation inferred from the dispersion relation with 3 subtractions.

 \subsection{Improved positivity bounds}

Focussing now on the scattering configuration \RNum{1}, the improved positivity bound is (expanding the integrals to NNLO in powers of $m/(\epsilon \Lambda_c)$),
\be
\begin{split}
0<&\,\partial_s^2 \tilde{\mathcal{A}}^{\textrm{\RNum{1}}}(0,0,0)-\frac{2}{\pi}\int_{0}^{\epsilon^2\Lambda_c^2}\rd s'\,\frac{\textrm{Disc}_{s}\mathcal{A}^{\textrm{\RNum{1}}}(s',0,u')}{s^{'3}}-\frac{2}{\pi}\int_{0}^{\epsilon^2\Lambda_c^2}\rd u'\,\frac{\textrm{Disc}_{u}\mathcal{A}^{\textrm{\RNum{1}}}(s',0,u')}{u^{'3}}\\
0<&\frac{11 e^4}{360\pi^2 m^4}-\frac{11e^2}{180\pi^2 m^2 \mpl^2}-\frac{2}{\pi}\int_{0}^{\epsilon^2\Lambda_c^2}\rd s'\,\frac{\textrm{Disc}_{s}\mathcal{A}^{\textrm{\RNum{1}}}(s',0,u')}{s^{'3}}-\frac{2}{\pi}\int_{0}^{\epsilon^2\Lambda_c^2}\rd u'\,\frac{\textrm{Disc}_{u}\mathcal{A}^{\textrm{\RNum{1}}}(s',0,u')}{u^{'3}}\\
0<&-\frac{11 e^2}{360 \pi ^2 m^2 \mpl^2}-\frac{e^2}{3 \pi ^2 \Lambda ^2 \mpl^2}-\frac{e^4}{4 \pi ^2 \Lambda ^4}-\frac{e^2 m^2}{4 \pi ^2 \Lambda ^4 \mpl^2}+\frac{e^4 }{\pi ^2 \Lambda ^4}\ln \frac{\Lambda}{m} +\frac{e^2 m^2 }{\pi ^2 \Lambda ^4 \mpl^2} \ln \frac{\Lambda}{m}\,,
\end{split}
\ee
where $\Lambda=\epsilon\Lambda_c$. Once again focusing on the EFT hierarchy $m \ll \epsilon \Lambda_c \ll \mpl$ this is effectively
\be
\frac{e^4 }{\pi ^2 \Lambda ^4} \left( \ln \frac{\Lambda}{m} -\frac{1}{4} \right) -\frac{11 e^2}{360 \pi ^2 m^2 \mpl^2}>0\,.
\ee
 As before, without gravity the bound is trivially satisfied, in the presence of gravity the bound is violated if the cutoff is taken to infinity. If however the cutoff scale is taken below
 \be\label{lowcutoff}
\epsilon\Lambda_c \lesssim (e m\mpl)^{1/2}\,,
\ee
positivity is respected. Up to numerical factors this is essentially the same order as the bound derived in scalar QED and suggests a universal result.

\subsection{Higher order gravitational contributions}\label{higherorder}

\begin{figure}[t]
    \centering
\includegraphics[width=10cm]{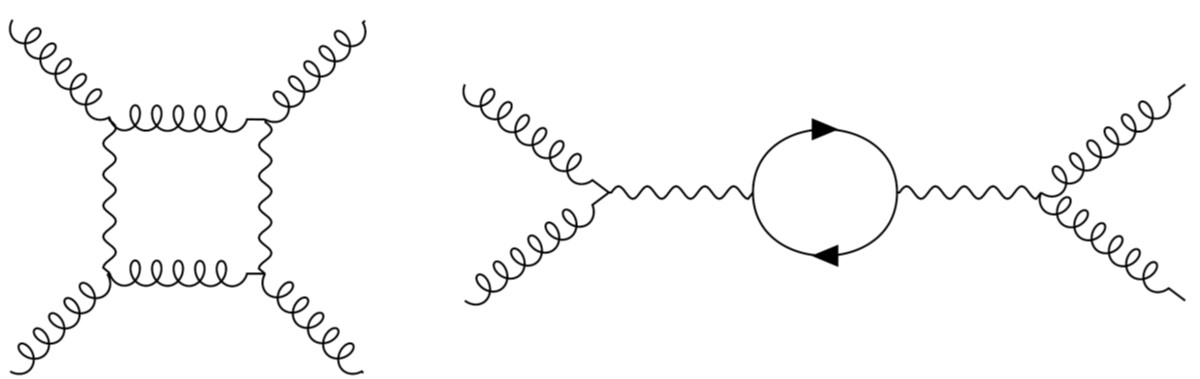}
\caption{Example gravitational contributions at order $1/\mpl^4$. }
\label{fig:higherorder}
\end{figure}

Up to now, we have  only considered the gravitational corrections to scattering amplitudes up to order $1/\mpl^2$ on the grounds that these dominate. Given the results of the improved positivity bounds, it would be remiss not to address whether higher order corrections, specifically the next ones at order $1/\mpl^4$ could rescue positivity without the need for the low cutoff \eqref{lowcutoff}. Indeed as noted already in \cite{Cheung:2014ega} this is in principle possible. In the context of the improved positivity bounds derived here we can show that this actually leads to equally strong implications.
Example Feynman diagrams at this order are given in Fig.~\ref{fig:higherorder}. They contain no electromagnetic vertices and therefore do not vanish as $e \rightarrow 0 $. Indeed any matter species, even uncharged, will give similar contributions. Furthermore these amplitudes are logarithmically divergent within 4D QED minimally coupled to gravity, necessitating the need to add curvature square operators in the actions \eqref{qed} and \eqref{sqed} whose coefficients can only be determined by matching onto an unknown UV completion.
When included in the amplitude, the improved positivity bounds for spinor QED become
\be\label{higherorderimproved}
\frac{e^4 }{\pi ^2 \Lambda ^4} \left( \ln \frac{\Lambda}{m} -\frac{1}{4} \right) -\frac{11 e^2}{360 \pi ^2 m^2 \mpl^2} - \frac{B}{\mpl^4} \ln \left(\frac{\Lambda}{m} \right) + \frac{\gamma_{m}}{\mpl^4}>0\,,
\ee
with a similar expression for scalar QED. Here $B$ is a known positive $\mathcal{O}(1)$ coefficient determined from the positive discontinuities of the diagrams in Fig.~\ref{fig:higherorder}, and $\gamma_{m}$ is an unknown matching coefficient accounting for the curvature square types of operators which needed to be added to the actions \eqref{qed} defined for convenience\footnote{More generally $\gamma(\mu)= \gamma_{m}- B \ln(\mu/m)$ demonstrating the logarithmic running of the diagrams in Fig.~\ref{fig:higherorder}.} at a fixed RG scale $\mu \sim m$. The logarithmic $\Lambda$ dependence of the $B$ term arises, as in the first term, from the application of the improved bounds which removes the branch cut up to the scale $s'\sim\Lambda^2$. We now see that in principle there is another solution to maintain positivity, other than that of \eqref{lowcutoff}. Indeed assuming $\Lambda \gg (e m\mpl)^{1/2}$ then \eqref{higherorderimproved} effectively becomes
\be\label{higherorderimproved2}
-\frac{11 e^2}{360 \pi ^2 m^2 \mpl^2}+ \frac{\gamma_{\Lambda}}{\mpl^4}>0\,.
\ee
Now if $\gamma_{\Lambda}$ is order unity, then \eqref{higherorderimproved2} amounts to a bound of the form $m \gtrsim e \mpl$, in complete opposition to what is anticipated from the weak gravity conjecture. This is of course because we are trying to maintain positivity with terms which are higher order in $1/\mpl$ rather than lower order. A similar conclusion was made in the 3D case in \cite{Chen:2019qvr}. Unlike the situation in 3D however, we cannot add additional uncharged light states to remove this tension. That is because in 3D, the $R^2$ terms do not need renormalization and any matter fields, even uncharged, contribute to them as $\Delta S \sim \int \d^3 x \sqrt{-g} R^2/m$. Thus by including very light uncharged fields, such as the neutrino, we can maintain overall positivity without needing to satisfy $m \gtrsim e \mpl$. \\

Returning to four dimensions, more generally we should account for the role of larger $\gamma_{\Lambda}$ which cannot be determined within the QED EFT. The typical expectation for the magnitude of $\gamma_{\Lambda}$ is of order the number of fields $N_{*}$ that lie below the Planck scale since every matter field contributes a term of this form on integrating out. Then the improved positivity bound \eqref{higherorderimproved2} can be satisfied provided
\be \label{higherorderimproved3}
m \gtrsim  \frac{e\mpl}{\sqrt{N_*}}\, .
\ee
In particular, for a weakly coupled UV completion in which new massive spin $2$ and higher states arise at a scale $M_s$, for which $\mpl^2 = M_s^2/g_s^2$, then the scale expected for $\gamma_{\rm UV}$ is $\gamma_{\rm UV} \sim \mpl^2/M_s^2 \sim 1/g_s^2 \gg 1$. In this case \eqref{higherorderimproved2}  amounts to
\be  \label{higherorderimproved4}
m  \gtrsim e M_s \, .
\ee
Unless $e$ is extremely small for {\it every} charged states in the theory, both of the bounds \eqref{higherorderimproved3} and \eqref{higherorderimproved4} are unreasonable constraints on theories of interest, and so we do not consider this `resolution' to maintain positivity as a meaningful solution. Furthermore they stand in clear opposition to the expectations from weak gravity conjecture \cite{ArkaniHamed:2006dz}.

\section{Discussion}

In this article we have considered whether QED minimally coupled to gravity respects positivity bounds applied with the $t$-channel pole removed. Regardless of whether we consider charged fermions or scalars, we find that it only does so if the effective field theory itself breaks down at the low scale $\Lambda_{\rm new} \sim (e m \mpl)^{1/2}$, $m$ being the mass of the electron. This result was already anticipated in the renormalizable scalar field theories discussed in \cite{Alberte:2020jsk}, and we see that the new features of gauge invariance and spin do not change the essential implications. Furthermore, these results are easily generalized to $N_f$ spinors and $N_s$ scalars given that the entire effect comes from one-loop diagrams in which the matter (i.e. electron) is in the loop, and so the relevant amplitude contributions are proportional to $N_f$ and $N_s$ respectively. Crucially since both scalars and spinors give a characteristically negative contribution to the positivity bound at order $1/\mpl^2$ in graviton exchange then no choice of $N_f$ and $N_s$ can be used to cancel these contributions and affect these conclusions. \\

There are three possible perspectives we may take on these results:
\begin{itemize}
\item {\it Either} consistent (local) UV completions of QED coupled to gravity do require new physics at scale $\Lambda_{\rm new} \sim (e m \mpl)^{1/2}$, regardless of whether the UV completion is weakly coupled or strong coupled,
\item {\it Or} for every charged state, we must impose the unreasonable bounds $m \gtrsim e \mpl/\sqrt{N_*}$, where $N_*$ is the number of fields below the Planck scale, as discussed in section~\ref{higherorder},
\item {\it Or} the positivity bounds do not apply to the $t$-channel pole subtracted amplitude.
\end{itemize}
The first conclusion is remarkable in that it is far more stringent than the cutoff expected from the weak gravity conjecture, namely $\Lambda_{\rm WG} \sim e \mpl$ \cite{ArkaniHamed:2006dz}, and by extension it is lower than the scale $e^{1/3} \mpl$ \cite{Heidenreich:2016aqi,Heidenreich:2017sim} which is obtained with assumptions on the UV completion from the species bound and Landau pole. Thus if taken seriously,  in this context positivity bounds are far more constraining than other `swampland' considerations.
The second option while technically valid is a rather unreasonable condition for theories in which the electric charge of all the states is not incredibly small, as in the case of real QED, and so we do not consider it further. The last possibility was discussed in more detail in \cite{Alberte:2020jsk} where we noted there are several reasons to doubt strict positivity in the gravitational context.

\paragraph{\bf Absence of decoupling limit:} As discussed in \cite{Alberte:2020jsk}, the only case in which one can be sure that positivity holds is when their is a clear $\mpl \rightarrow \infty$ decoupling limit for which the $t$-channel pole drops out, provided other terms in the amplitude do not vanish. Interestingly, we can see that this is not possible here without introducing other problems. For instance, for standard QED, $N_f=1$, in order to take a decoupling limit $\mpl \rightarrow \infty$ keeping the scale $\Lambda_{\rm new} \sim (e m \mpl)^{1/2}$ at least constant for fixed $m$, or indeed the weak gravity conjecture scale $\Lambda_{\rm WG} = e \mpl$ at least constant, we would need to scale $e \sim 1/\mpl$ meaning we send $\mpl \rightarrow \infty$ for fixed $\Lambda_{\rm WG}$ or $\Lambda_{\rm new}$. In doing so the non-gravitational part of the amplitude which is of order $e^4/m^4 \sim 1/\mpl^4$ vanishes faster than the $\mpl^2$ suppressed $t$-channel pole, undermining the very purpose of the decoupling limit since then clearly the graviton exchange dominates. Similarly for the improved amplitude \eqref{improvedamplitude} the non-gravitational part scales as $e^4/\Lambda_{\rm new}^4\sim 1/\mpl^4$ and we reach the same conclusion. \\

In many cases a better decoupling limit is obtained by taking $N=N_f+N_s$ species and a large $N$ limit. For instance, one may be tempted to consider a theory with $N_f$ fermions, so that the $e^4$ suppression of the one-loop amplitude can be compensated by scaling $N_f$ faster than $\mpl^2$. In such a limit $N_f e^4/m^4$  and the one-loop gravitational corrections of the form $N_f e^2/(m^2 \mpl^2)$  dominate over the $t$-channel pole, suggesting at first sight that an appropriate $\mpl\to \infty$ decoupling limit could be achieving while maintaining some of the relevant physical implications. However doing so necessarily runs into problems with the species bound \cite{Dvali:2007hz,Dvali:2007wp} since then $\Lambda_{\rm species} \sim \mpl/\sqrt{N_f} \rightarrow 0$. While not constituting a proof, these arguments are highly suggestive that we should not enforce strict positivity, given the absence of a clean decoupling limit, but rather a weaker condition of the form \eqref{newpositivitiy}, or more precisely in the present context \eqref{newimprovedpositivity}. We are of course free to take the decoupling limit $\mpl \rightarrow \infty$ for fixed $e$ and $m$ which in the string amplitude context amounts to $g_s \rightarrow 0$ so that the amplitudes are dominated by tree level contributions. However in this case $\Lambda_{\rm new} \rightarrow \infty$ and no contradiction is observed from applying positivity bounds to the tree amplitudes in the non-gravitational limit. It is clear that to make further progress it is crucial to establish to what extent positivity bounds apply with gravity, specifically whether it is in the weak sense\footnote{A related discussion is given in the arXiv:v4 version of \cite{Bellazzini:2019xts}, where the weakening is attributed to a violation of Froissart on the UV. Our perspective is that this assumption is not necessary.} \eqref{newpositivitiy} or \eqref{newimprovedpositivity}, or the stronger one $c>0$ utilized here.

\bigskip
\noindent{\textbf{Acknowledgments:}}
 The work of AJT and CdR is supported by STFC grants ST/P000762/1 and ST/T000791/1. CdR thanks the Royal Society for support at ICL through a Wolfson Research Merit Award. LA and CdR are supported by the European Union's Horizon 2020 Research Council grant 724659 MassiveCosmo ERC--2016--COG. CdR is also supported by a Simons Foundation award ID 555326 under the Simons Foundation's Origins of the Universe initiative, `\textit{Cosmology Beyond Einstein's Theory}' and by a Simons Investigator award 690508.  SJ is supported by an STFC studentship. AJT thanks the Royal Society for support at ICL through a Wolfson Research Merit Award.

\appendix



\section{Scalar QED}\label{1loopscalarqed}

\subsection{Conventions}
\label{sec:conventions}
Before jumping into the core of the derivations, it is useful to summarize our relevant conventions. We parameterize the physical momenta of the four particles as
\be
k^\mu_i=(k,k\sin\vartheta_i,0,k\cos\vartheta_i)\,,
\ee
with
\be
\vartheta_1=0,\quad\vartheta_2=\pi,\quad\vartheta_3=\theta,\quad\vartheta_4=\pi+\theta\,,
\ee
where the particles $1,2$ are ingoing and $3,4$ are outgoing. The quantities $k$ and $\theta$ are expressed through the Mandelstam variables using the relations $k^2=s/4$ and $\cos\theta=1+2t/s$. Similarly, the normalized and transverse polarization vectors are defined as
\be
\varepsilon_i^\mu(\pm)=\frac{1}{\sqrt{2}}\left(0,\cos\vartheta_i,\pm i,-\sin\vartheta_i\right)\,.
\ee
The $2\to 2$ scattering amplitude in conventions where the momenta are associated with indices as $A^\mu(k_1)$, $A^\nu(k_2)$, $A^\alpha(k_3)$, $A^\beta(k_4)$ reads
\be
\mathcal A_{\rm physical}(h_1,h_2,h_3,h_4)=\varepsilon^\mu_1(h_1)\varepsilon^\nu_2(h_2)\mathcal A_{\rm physical}^{\mu\nu\alpha\beta} (k_1,k_2,k_3,k_4)\varepsilon_3^{*\alpha}(h_3)\varepsilon_4^{*\beta}(h_4)\,,
\ee
where $h_i=\pm1$ are the helicities of each particle.

\paragraph{All-ingoing notations.} Henceforth, for computational simplicity we treat also the particles $3,4$ as ingoing by reversing their four-momenta: $k_3^\mu=-(k,k\sin\theta,0,k\cos\theta)$ and $k_4^\mu=-(k,-k\sin\theta,0,-k\cos\theta)$. We leave the polarization vectors unchanged and calculate the $2\to2$ scattering amplitude as
\be\label{allingoing}
\mathcal A_{\rm ingoing}(h_1,h_2,h_3,h_4)=\varepsilon^\mu_1(h_1)\varepsilon^\nu_2(h_2)\mathcal A_{\rm ingoing}^{\mu\nu\alpha\beta} (k_1,k_2,k_3,k_4)\varepsilon_3^{\alpha}(h_3)\varepsilon_4^{\beta}(h_4)\,.
\ee
Since under complex conjugation the helicity flips sign, i.e. $\varepsilon_i^{*\mu}(h_i)=\varepsilon_i^{\mu}(-h_i)$ then an all-ingoing amplitude is mapped to a physical amplitude as $\mathcal A_{\rm ingoing}(h_1,h_2,h_3,h_4)=\mathcal A_{\rm physical}(h_1,h_2,-h_3,-h_4)$. We only use the all-ingoing amplitude \eqref{allingoing} throughout the text and drop the subscript from now on.

We define $\kep_{ij}\equiv k_{i}\cdot\varepsilon_{j}$ and $\vep_{ij}\equiv\varepsilon_{i}\cdot\varepsilon_{j}$. The inner products between external polarisations in terms of the Mandelstam variables are,
\be\begin{split}
\vep_{12}&=-\frac12-\frac{h_1 h_2}{2}\,,\quad\vep_{13}=-\frac{h_1 h_3}{2}+\frac12+\frac{t}{s}\,,\quad\vep_{14}=-\frac{h_1 h_4}{2}-\frac12-\frac{t}{s}\,,\\
\vep_{34}&=-\frac12-\frac{h_3 h_4}{2}\,,\quad\vep_{24}=-\frac{h_2 h_4}{2}+\frac12+\frac{t}{s}\,,\quad\vep_{23}=-\frac{h_2 h_3}{2}-\frac12-\frac{t}{s}\,.
\end{split}\ee
The inner products between external momenta and polarisations are,
\be
\kep_{13}=\kep_{24}=-\frac{\sqrt{t u}}{\sqrt{2} \sqrt{s}}\,,\quad\kep_{14}=\kep_{23}=\frac{\sqrt{t u}}{\sqrt{2} \sqrt{s}}\,.
\ee
These are symmetric $\kep_{ij}=\kep_{ji}$. All other inner products vanish. Note that if all helicities are flipped all inner products are unchanged since helicity dependence always appears in the form $h_i h_j$.

\subsection{Tree-level Photon-Graviton contributions}
\label{sec:Tree_level}

Irrespectively on whether we are interested in the scalar QED Lagrangian \eqref{sqed} or the spinor one \eqref{qed}, the tree-level photon-graviton contributions are the same. Writing the metric as $g\mn=\eta\mn+h\mn/\mpl$, the photon-photon-graviton interactions are,
\be
\mathcal{L}_{hAA}=\frac{1}{2 \mpl} h^{\mu
\nu}T\mn=\frac{1}{2 \mpl} h^{\mu
\nu}\( F_{\mu\alpha}F_\nu{}^\alpha-\frac 14 F^{\alpha\beta}F_{\alpha \beta}\eta\mn\)\,,
\ee
where all indices are raised/lowered with the Minkowski metric $\eta\mn$.
The Feynman rule for the $h^{\alpha\beta}A^{\mu}(k_1)A^{\nu}(k_2)$ vertex is then,
\be\label{treeAAH}
\mathcal{V}^{\mu\nu;\alpha\beta}\equiv\frac{\ri}{2}k_1\cdot k_2\left(\eta^{\alpha\beta}\eta^{\mu\nu}-2\eta^{\nu(\beta}\eta^{\alpha)\mu}\right)-\frac{\ri}{2}\eta^{\alpha\beta}k_1^{\nu} k_2^{\mu}-\ri\eta^{\mu\nu}k_{1}^{(\alpha}k_{2}^{\beta)}+\ri\left( k_2^\mu \eta^{\nu(\alpha}k_1^{\beta)}+ k_2^{(\beta} \eta^{\alpha)\mu}k_1^\nu\right)\,.
\ee
The $d$-dimensional graviton propagator is (in harmonic/de Donder gauge),
\be
D_{\alpha\beta;\gamma\delta}(k^2)=-\frac{2 \ri}{k^2} \left(\eta_{\alpha\delta} \eta_{\beta\gamma}+\eta_{\alpha\gamma} \eta_{\beta\delta}-\frac{2 \eta_{\alpha\beta} \eta_{\gamma\delta}}{d-2}\right)\,.
\ee
At tree level we are free to set $d=4$. Then the $s$-diagram gives,
\be
\begin{split}
  \mpl^2  \mathcal{A}_{s}&=-\ri\vep_{1\chi}\vep_{2\kappa}\mathcal{V}^{\chi\kappa;\alpha\beta}D_{\alpha\beta;\gamma\delta}(-s)\mathcal{V}^{\mu\nu,\gamma\delta}\vep_{3\mu}\vep_{4\nu}\\
    &=\vep_{12}\vep_{34}\left(s-\frac{t^2}{2s}-\frac{u^2}{2s}\right)-\frac12(\vep_{14}\vep_{23}+\vep_{13}\vep_{24})s+\frac{tu}{s}\left(\vep_{12}+\vep_{34}\right)\\
    &=\frac{-1}{8s}\Big(t^2 (h_1 (h_2 h_3 h_4-h_2+h_3-h_4)-h_2 (h_3- h_4)-h_3 h_4+1)\\&\qquad\quad+u^2 (h_1 (h_2 h_3 h_4-h_2-h_3+h_4)-h_4 (h_2+h_3)+h_2 h_3+1)\Big)\,.
\end{split}
\ee
The $t$-diagram gives,
\be
\begin{split}
   \mpl^2 \mathcal{A}_{t}&=-\ri\vep_{1\chi}\vep_{3\kappa}\mathcal{V}^{\chi\kappa;\alpha\beta}D_{\alpha\beta;\gamma\delta}(-t)\mathcal{V}^{\mu\nu,\gamma\delta}\vep_{2\mu}\vep_{4\nu}\\
    &=\frac{-1}{8t}\Big(u^2 (h_1 (h_2 h_3 h_4-h_2-h_3+h_4)-h_4 (h_2+h_3)+h_2 h_3+1)\\&\qquad\quad+s^2 (h_1 (h_2 h_3 h_4+h_2-h_3-h_4)-h_2 (h_3+h_4)+h_3 h_4+1)\Big)\,,
\end{split}
\ee
and the $u$-diagram gives,
\be
\begin{split}
  \mpl^2  \mathcal{A}_{u}&=-\ri\vep_{1\chi}\vep_{4\kappa}\mathcal{V}^{\chi\kappa;\alpha\beta}D_{\alpha\beta;\gamma\delta}(-u)\mathcal{V}^{\mu\nu,\gamma\delta}\vep_{2\mu}\vep_{3\nu}\\
    &=\frac{-1}{8u}\Big(t^2 (h_1 (h_2 h_3 h_4-h_2+h_3-h_4)+h_2 (h_4-h_3)-h_3 h_4+1)\\&\qquad\quad+s^2 (h_1 (h_2 h_3 h_4+h_2-h_3-h_4)-h_2 (h_3+h_4)+h_3 h_4+1)\Big)\,.
\end{split}
\ee
Combining those three channels together we have
\be
\begin{split}
\mathcal A_{\rm tree, 0}(h_1,h_2,h_3,h_4)=\frac{-1}{4\mpl^2stu}&\left[(-1+h_1h_2)(-1+h_3h_4)\left(t^4+u^4-\frac{1}{4}(s^2+t^2+u^2)^2\right)\right.\\
+&\,\,\,(-1+h_1h_3)(-1+h_2h_4)\left(s^4+u^4-\frac{1}{4}(s^2+t^2+u^2)^2\right)\\
-&\,\,\left.(-1+h_2h_3)(-1+h_1h_4)\left(u^4-\frac{1}{4}(s^2+t^2+u^2)^2\right)\right]\,,
\end{split}
\ee
which give the familiar results for example processes
\be
\begin{split}
\mathcal{A}(++++)&=0 =\mathcal{A}(+++-)\, , \\
\mathcal{A}(++--)&=\frac{1}{\mpl^2}\frac{s^{4}}{stu}\,, \\
\mathcal{A}(+-+-)&=\frac{1}{\mpl^2}\frac{t^{4}}{stu}\,, \\
\mathcal{A}(+--+)&=\frac{1}{\mpl^2}\frac{u^{4}}{stu}\,.
\end{split}
\ee

\subsection{Photon wavefunction renormalization}
\begin{figure}[t]
\centering
\begin{subfigure}{0.45\textwidth}
\centering
\includegraphics[height=2cm]{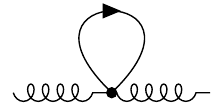}
\caption{Specific to scalar QED.} \label{fig:wfr1}
\end{subfigure}
\hspace*{0.3in} 
\begin{subfigure}{0.45\textwidth}
\centering
\includegraphics[height=2cm]{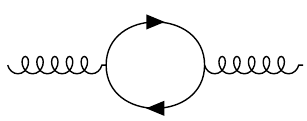}
\caption{For both scalar and spinor QED.} \label{fig:wfr2}
\end{subfigure}
\caption{One loop self-energy contribution for scalar and spinor QED.}
\label{fig:wfr}
\end{figure}

At one-loop there are two diagrams contributing to the quantum photon propagator given in Fig.~\ref{fig:wfr}.
The self-energy of the photon $\Pi^{\rho\sigma}(k^2)$ obeys the Ward identity $k_{\rho}\Pi^{\rho\sigma}(k^2)=0$ implying that the self-energy is proportional to the projector onto the subspace transverse to $k^{\mu}$,
\be
\Pi^{\rho\sigma}(k^2)=\left(k^{2}\eta^{\rho\sigma}-k^{\rho}k^{\sigma}\right)\Pi(k^2)\,.
\ee
The two diagrams give,
\be
\begin{split}
    \Pi^{\rho\sigma}(k^2)&=e^2\,\mu^\epsilon\int\ddp\,\frac{(2p-k)^\rho(2p-k)^\sigma}{(p^2+m^2)((p-k)^2+m^2)}-2e^2\eta^{\rho\sigma}\int\ddp\,\frac{1}{p^2+m^2}\\
    &=e^2\,\mu^\epsilon\int\ddp\,\frac{(2p-k)^\rho(2p-k)^\sigma-2\eta^{\rho\sigma}((p-k)^2+m^2)}{(p^2+m^2)((p-k)^2+m^2)}\\
    &=e^2\,\mu^\epsilon\int\ddl\int_0^1\rd x\,\frac{\left(\frac{4}{d}-2\right)\eta^{\rho\sigma}l^2-2\eta^{\rho\sigma}(k^2+m^2+k^2x^2-2k^2x)+k^\rho k^\sigma(1-4x+4x^2)}{(l^2+m^2+xk^2(1-x))^2}\\
    &=-\frac{\ri e^2}{4\pi^2}\int_{-\frac12}^{\frac12}\rd y\,y^{2}(k^2\eta^{\rho\sigma}-k^\rho k ^\sigma)\ln{\frac{\mu^2}{m^2-k^2(y^2-\frac{1}{4})}}\,,
\end{split}
\ee
where in the last step the Feynman parameter is redefined by $y=x-\frac{1}{2}$. By re-summing the series of 1PI diagrams contributing to the quantum propagator, one can show that the wavefunction renormalisation is,
\be
Z_{A}=\frac{1}{1+\ri\Pi(0)}\,.
\ee
The scalar part of the self-energy can be found from the expression above to be
\be
\Pi(k^2)=-\frac{\ri e^2}{4\pi^2}\int_{-\frac12}^{\frac12}\rd y\,y^{2}\ln{\frac{\mu^2}{m^2-k^2(y^2-\frac{1}{4})}}\,,
\ee
so that,
\be
\Pi(0)=-\frac{\ri e^2}{48\pi^2}\ln{\frac{\mu^2}{m^2}}\,,
\ee
and thus,
\be
Z_{A}=1-\frac{e^2}{48\pi^2}\ln{\frac{\mu^2}{m^2}}+\mathcal{O}(e^4)\,.
\ee

\subsection{Non-gravitational Contributions}\label{scalarQEDnongrav}

By non-gravitational contributions, we refer to the Feynman diagrams with no graviton lines, shown in the first line of Fig.~\ref{fig:scattering2}. There are no non-gravitational tree diagrams so we start at one-loop. From diagrams with 2 internal propagators we get three crossing related diagrams giving the familiar expressions,
\be
\begin{split}
    \mathcal{A}_{2-s}&=\frac{e^4}{4\pi^2}(\varepsilon_1\cdot\varepsilon_2)(\varepsilon_3\cdot\varepsilon_4)\int_{0}^{1}\rd x\,\left(\msbar+\ln{\frac{\mu^{2}}{m^{2}+sx(x-1)}}\right) \, , \\
    \mathcal{A}_{2-t}&=\frac{e^4}{4\pi^2}(\varepsilon_1\cdot\varepsilon_3)(\varepsilon_2\cdot\varepsilon_4)\int_{0}^{1}\rd x\,\left(\msbar+\ln{\frac{\mu^{2}}{m^{2}+tx(x-1)}}\right)\, , \\
    \mathcal{A}_{2-u}&=\frac{e^4}{4\pi^2}(\varepsilon_1\cdot\varepsilon_4)(\varepsilon_2\cdot\varepsilon_3)\int_{0}^{1}\rd x\,\left(\msbar+\ln{\frac{\mu^{2}}{m^{2}+ux(x-1)}}\right)\,.
\end{split}
\ee
Next we have diagrams that have three internal $\phi$ propagators in the loop which give (including a factor of 2 for charge flow reversal),
\be
\begin{split}
\mathcal{A}_{3-s}&=\ri32e^4\mu^{\epsilon}\int\ddl\rd x\,\rd y\,\frac{\frac{2}{d}l^2\,\varepsilon_{12}\,\varepsilon_{34}-\kep_{12}\kep_{21}\,\varepsilon_{34}\, x y-\kep_{34}\,\kep_{43}\,\vep_{12}\, x y}{(l^2+m^2-sxy)^3}\\
&=-\frac{e^4}{\pi^2}\int\rd x\,\rd y\,\left(\vep_{12}\vep_{34}\ln{\frac{\mu^2}{m^2-sxy}}-\frac{xy\left(\kep_{12}\kep_{21}\vep_{34}+\kep_{34}\kep_{43}\vep_{12}\right)}{\left(m^2-sxy\right)}\right)\\
&=-\frac{e^4}{\pi^2}\int\rd x\,\rd y\,\left(\vep_{12}\vep_{34}\ln{\frac{\mu^2}{m^2-sxy}}\right)\,.
\end{split}
\ee
The other channels are given by,
\be
\begin{split}
\mathcal{A}_{3-t}&=-\frac{e^4}{\pi^2}\int\rd x\,\rd y\,\left(\vep_{13}\vep_{24}\ln{\frac{\mu^2}{m^2-txy}}-\frac{xytu\left(\vep_{24}+\vep_{13}\right)}{2s\left(m^2-txy\right)}\right)\,,\\
\mathcal{A}_{3-u}&=-\frac{e^4}{\pi^2}\int\rd x\,\rd y\,\left(\vep_{14}\vep_{23}\ln{\frac{\mu^2}{m^2-uxy}}-\frac{xytu\left(\vep_{23}+\vep_{14}\right)}{2s\left(m^2-uxy\right)}\right)\,.
\end{split}
\ee
Finally there are three diagrams (in addition to their  charge reversal) with four internal $\phi$ propagators, i.e. the `box' diagram, as the last diagram on the first line of Fig.~\ref{fig:scattering2}
\be
\ri\mathcal{A}_{4-1}=16e^4\mu^{\epsilon}\int\ddp\frac{(p\vep_1)(p\vep_3)(p\vep_4+\kep_{34})(p\vep_2-\kep_{12})}{(p^2+m^2)((p+k_3)^2+m^2)((p-k_1)^2+m^2)((p+(k_3+k_4))^2+m^2)}\,.
\ee
Shifting the loop momentum, $l=p+xk_3-yk_1+z(k_3+k_4)$ and including the factors of 2 from charge reversal,
\be
\ri\mathcal{A}_{4-s}=2\times3!\times16e^4\mu^{\epsilon}\int\ddl\,\rd x\,\rd y\,\rd z\,\frac{A\,l^4+B\,l^2+C}{\left(l^2+m^2+sz(x+z+y-1)-tyx\right)^4} \, ,
\ee
where,
\be\begin{split}
    A&=\frac{\vep_{12}\vep_{34}+\vep_{13}\vep_{24}+\vep_{14}\vep_{23}}{d(d+2)}\,,\\
    B&=\frac{-tu(\vep_{34}x^2+(\vep_{13}-\vep_{23}-\vep_{14}+\vep_{24})xy+\vep_{12}y^2)}{2ds}\,,\\
    C&=\frac{t^2u^2x^2y^2}{4s^2}\,.
\end{split}
\ee
The other diagrams are computed similarly,
\be
\ri\mathcal{A}_{4-u}=2\times3!\times16e^4\mu^{\epsilon}\int\ddl\,\rd x\,\rd y\,\rd z\,\frac{A\,l^4+D\,l^2+E}{\left(l^2+m^2+uz(x+z+y-1)-tyx\right)^4} \, ,
\ee
where,
\be\begin{split}
    D&=\frac{tu}{2ds^2}f\left(\{h_i\},s,t\right)\,,\\
    E&=\frac{t^2 u^2 (x+z-1) (x+z) (y+z-1) (y+z)}{4 s^2}\,,
\end{split}
\ee
where $f$ is linear in $s$ and $t$ (with no terms like $st$).
Finally the $t$-channel contribution gives
\be
\ri\mathcal{A}_{4-t}=2\times3!\times16e^4\mu^{\epsilon}\int\ddl\,\rd x\,\rd y\,\rd z\,\frac{A\,l^4+F\,l^2+C}{\left(l^2+m^2+sz(x+z+y-1)-uyx\right)^4} \, ,
\ee
where,
\be\begin{split}
    F&=\frac{-tu(\vep_{34}x^2-(\vep_{13}-\vep_{23}-\vep_{14}+\vep_{24})xy+\vep_{12}y^2)}{2ds}\,.
\end{split}
\ee
The full expression for $D$ is,
\be
\begin{split}
    D=&\frac{h_1 h_2 t u y z}{2 d s}+\frac{h_1 h_2 t u y^2}{4 d s}-\frac{h_1 h_2 t u y}{4 d s}+\frac{h_1 h_2 t u z^2}{4 d s}-\frac{h_1 h_2 t u z}{4 d s}+\frac{h_1 h_3 t u x y}{4 d s}+\frac{h_1 h_3 t u x z}{4 d s}\\
    &-\frac{h_1 h_3 t u x}{4 d s}+\frac{h_1 h_3 t u y z}{4 d s}-\frac{h_1 h_3 t u y}{4 d s}+\frac{h_1 h_3 t u z^2}{4 d s}-\frac{h_1 h_3 t u z}{2 d s}+\frac{h_1 h_3 t u}{4 d s}-\frac{h_1 h_4 t u x y}{4 d s}\\
    &-\frac{h_1 h_4 t u x z}{4 d s}-\frac{h_1 h_4 t u y z}{4 d s}+\frac{h_1 h_4 t u y}{4 d s}+\frac{h_1 h_4 t u z}{4 d s}-\frac{h_1 h_4 t u z^2}{4 d s}-\frac{h_2 h_3 t u x y}{4 d s}-\frac{h_2 h_3 t u x z}{4 d s}\\
    &+\frac{h_2 h_3 t u x}{4 d s}-\frac{h_2 h_3 t u y z}{4 d s}+\frac{h_2 h_3 t u z}{4 d s}-\frac{h_2 h_3 t u z^2}{4 d s}+\frac{h_2 h_4 t u x y}{4 d s}+\frac{h_2 h_4 t u x z}{4 d s}+\frac{h_2 h_4 t u y z}{4 d s}\\
    &+\frac{h_2 h_4 t u z^2}{4 d s}+\frac{h_3 h_4 t u x z}{2 d s}+\frac{h_3 h_4 t u x^2}{4 d s}-\frac{h_3 h_4 t u x}{4 d s}+\frac{h_3 h_4 t u z^2}{4 d s}-\frac{h_3 h_4 t u z}{4 d s}-\frac{2 t^2 u x y}{d s^2}\\
    &-\frac{t u x y}{d s}-\frac{2 t^2 u x z}{d s^2}-\frac{t u x z}{2 d s}+\frac{t^2 u x}{d s^2}+\frac{t u x^2}{4 d s}+\frac{t u x}{4 d s}-\frac{2 t^2 u y z}{d s^2}-\frac{t u y z}{2 d s}+\frac{t^2 u y}{d s^2}+\frac{t u y^2}{4 d s}\\
    &+\frac{t u y}{4 d s}+\frac{2 t^2 u z}{d s^2}-\frac{2 t^2 u z^2}{d s^2}+\frac{t u z}{2 d s}-\frac{t u z^2}{2 d s}-\frac{t^2 u}{2 d s^2}-\frac{t u}{4 d s}\,.
\end{split}
\ee
\paragraph{RG scale independence of non-gravitational contributions.}
A consistency check of numerical factors is to see that all dependence on the scale $\mu$ drops out of the total amplitude. The 2-propagator loop diagrams give the $\mu$ dependent term,
\be
\frac{e^4}{4\pi^2}\left(\vep_{12}\vep_{34}+\vep_{13}\vep_{24}+\vep_{14}\vep_{23}\right)\ln{\frac{\mu^2}{m^2}}\,,
\ee
from the 3-propagator diagrams we have,
\be
-\frac12\times\frac{e^4}{\pi^2}(\vep_{12}\vep_{34}+\vep_{13}\vep_{24}+\vep_{14}\vep_{23})\ln{\frac{\mu^2}{m^2}}\,,
\ee
from the 4-propagator diagrams we have,
\be
3\times2\times3!\times16\times\frac{1}{6}\times\frac{1}{384\pi^2}(\vep_{12}\vep_{34}+\vep_{13}\vep_{24}+\vep_{14}\vep_{23})\ln{\frac{\mu^2}{m^2}}\,.
\ee
Summing these terms gives zero so there is no $\mu$ dependence.


\subsection{One-loop graviton exchange}\label{ScalarQEDoneloop}
\begin{figure}[t]
    \centering
\includegraphics[width=13cm]{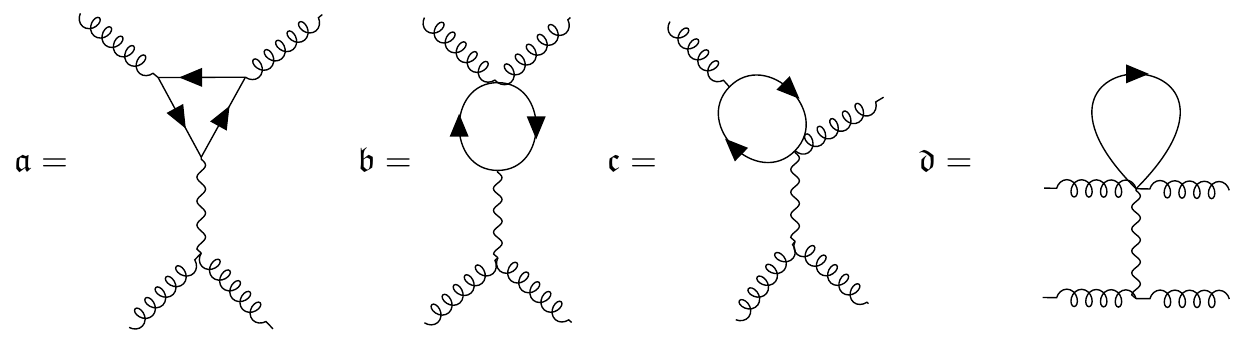}
\caption{Four classes of gravitational diagrams contributing at one-loop and to order $1/\mpl^2$. The arrows depict the direction of the charge flow. We do not show all the crossed versions of the diagrams. }
\label{fig:scattering4}
\end{figure}

The diagrams relevant for the one-loop graviton exchange are given in Fig.~\ref{fig:scattering4}. It is understood that each type of diagrams should also include flipped versions, with loops on the other end of the graviton line or on different photon lines as well as the crossed versions of each diagram. We now proceed with deriving the contributions to the discontinuity from each ones of these types of diagrams.

\subsubsection*{Type $\mathfrak{a}$ diagrams}
Summing the diagrams of type $\mathfrak{a}$ and their flipped version we obtain for the $t$-channel amplitude,
\be
\ri\mathcal{A}_{\mathfrak{a}-t}=(-\ri)^5\times2!\times2\times4e^2\int\ddl\,\rd x\,\rd y\,\frac{A\,l^4+B\,l^2+C}{\left(l^2+m^2-xyt\right)^3} \, .
\ee
The factor of 2! is due to combining propagators, the factor of 2 is due to doubling of the term by loop charge direction reversal and the factor of 4 just comes from the two QED like vertices. If we call the $(+,+,-,-),(+,-,-,+),(+,-,+,-)$ configurations \RNum{1},\RNum{2},\RNum{3}, respectively then the numerator coefficients are,
\begin{eqnarray}
    A_{\text{\RNum{1}}}&=&\frac{2 \ri \left(2 (d-2) s t+d s u+(d-4) s^2+(d-4) t (t+u)\right)}{d (d+2) s^2}=-\frac{8\ri}{d(d+2)}=-\frac{\ri}{3}-\frac{5 \ri \epsilon }{36}+\O(\epsilon^2)\nn\\
    A_{\text{\RNum{2}}}&=&\frac{2 \ri (d-4) (s+t) (s+t+u)}{d (d+2) s^2}=0\\
    A_{\text{\RNum{3}}}&=&\frac{2 \ri (d-4) t (t+u)}{d (d+2) s^2}=-\frac{2 \ri (d-4) t}{d (d+2) s}=\frac{\ri t\epsilon}{12s}+\O(\epsilon^2)\nn\\
    B_{\text{\RNum{1}}}&=&-\frac{2 \ri (s+t) (x+y-1) (s (x+y-1)+3 t (x+y)-t)}{d t}\nn\\
    B_{\text{\RNum{2}}}&=&-\frac{2 \ri (s+t)^2 (x+y-1)^2}{d t}\nn\\
    B_{\text{\RNum{3}}}&=&-\frac{\ri (d-4) t \left((d-2) x y (s+t)+2 m^2\right)}{(d-2) d s}=\frac{\ri t \epsilon  \left(m^2+s x y+t x y\right)}{4 s}+\O(\epsilon^2)\nn\\
    C_{\text{\RNum{1}}}&=&-\ri x y u^2 (x+y-1)^2 \, , \quad
    C_{\text{\RNum{2}}}=-\ri x y u^2 (x+y-1)^2 \, , \quad
     C_{\text{\RNum{3}}}=\frac{\ri (d-4) m^2 t x y u}{(d-2) s}=\mathcal{O}(\epsilon)\,.\nn
\end{eqnarray}
Note that for the $C$ coefficients, if they are of order $\epsilon$ they can be set to zero as the dim-reg integrals do not produce $1/\epsilon$ terms that would combine with them to make them finite. Moving on to the $s$-channel of diagram $\mathfrak{a}$ we have,
\be
\ri\mathcal{A}_{\mathfrak{a}-s}=(-\ri)^5\times2!\times2\times4e^2\int\ddl\,\rd x\,\rd y\,\frac{D\,l^4+E\,l^2+F}{\left(l^2+m^2-xys\right)^3} \, ,
\ee
\be
\begin{split}
    D_{\text{\RNum{1}}}&=\frac{2 \ri (d-4)}{d (d+2)}=-\frac{\ri \epsilon}{12}  +\O(\epsilon^2)\,,\\
    D_{\text{\RNum{2}}}&=-\frac{8 \ri u^2}{d (d+2) s^2}=-\frac{5 \ri   u^2\epsilon}{36 s^2}-\frac{\ri u^2}{3 s^2} \, , \quad    D_{\text{\RNum{3}}}=-\frac{8 \ri t^2}{d (d+2) s^2}=-\frac{5 \ri t^2 \epsilon }{36 s^2}-\frac{\ri t^2}{3 s^2} \, ,\\
    E_{\text{\RNum{1}}}&=\frac{2 \ri (d-4) m^2}{(d-2) d}=-\frac{\ri m^2\epsilon}{4} \, , \quad  E_{\text{\RNum{2}}}=0 \, , \quad    E_{\text{\RNum{3}}}=0 \, ,\quad   F_{\text{\RNum{1}}}=0 \, , \quad
    F_{\text{\RNum{2}}}=0\,,\quad
    F_{\text{\RNum{3}}}=0\,.
\end{split}
\ee
Finally, the $u$-channel gives,
\be
\ri\mathcal{A}_{\mathfrak{a}-u}=(-\ri)^5\times2!\times2\times4e^2\int\ddl\,\rd x\,\rd y\,\frac{G\,l^4+H\,l^2+I}{\left(l^2+m^2-xyu\right)^3}\,,
\ee
with
\begin{eqnarray}
    G_{\text{\RNum{1}}}&=&-\frac{8 \ri}{d (d+2)}=-\frac{5 \ri \epsilon }{36}-\frac{\ri}{3}+\mathcal{O}(\epsilon^2) \, , \quad  G_{\text{\RNum{2}}}=-\frac{2 \ri (d-4) u}{d (d+2) s}=\frac{\ri \epsilon  u}{12 s} \,, \quad    G_{\text{\RNum{3}}}=0 \, , \\
    H_{\text{\RNum{1}}}&=&-\frac{2\ri t(x+y-1)(2s(x+y)+t(-1+3x+3y))}{du} \, , \quad    H_{\text{\RNum{2}}}=\frac{\ri(d-4)u((d-2)txy-2m^2)}{(d-2)ds} \, , \nn\\
    H_{\text{\RNum{3}}}&=&-\frac{2 \ri t^2 \left(x+y-1\right)^2}{du} \, , \quad    I_{\text{\RNum{1}}}=-\ri t^2 x y (x+y-1)^2 \, , \quad I_{\text{\RNum{2}}}=\frac{\ri(d-4)m^2 t u x y}{(d-2)s} \, ,\nn \\
    I_{\text{\RNum{3}}}&=&-\ri t^2 x y (x+y-1)^2\,.\nn
\end{eqnarray}
\subsubsection*{Type $\mathfrak{b}$ diagrams}
For type $\mathfrak{b}$ diagrams (including their version flipped up-down) we find the general expression,
\be
\ri\mathcal{A}_{\mathfrak{b}-t}=(-\ri)^3\times2e^2\int\ddl\,\rd x\,\frac{J\,l^2+K}{(l^2+m^2-x(1-x)t)^2} \, ,
\ee
\be
\begin{split}
    J_{\text{\RNum{1}}}&=J_{\text{\RNum{2}}}=0 \, , \quad    J_{\text{\RNum{3}}}=\frac{2 \ri (d-4) t (t+u)}{d s^2} \, , \\
    K_{\text{\RNum{1}}}&=K_{\text{\RNum{2}}}=0 \, , \quad    K_{\text{\RNum{3}}}=-\frac{2 \ri (d-4) m^2 t}{(d-2) s} \, .\\
\end{split}
\ee
The $s$-channel has the expressions,
\be
\ri\mathcal{A}_{\mathfrak{b}-s}=(-\ri)^3\times2e^2\int\ddl\,\rd x\,\frac{L\,l^2+M}{(l^2+m^2-x(1-x)s)^2} \, ,
\ee
\be
\begin{split}
    L_{\text{\RNum{1}}}&=\frac{2 \ri (d-4)}{d}\, , \quad  L_{\text{\RNum{2}}}=0 \, , \quad   L_{\text{\RNum{3}}}=0 \, , \\
    M_{\text{\RNum{1}}}&=\frac{2 \ri (d-4) m^2 }{d-2} \, , \quad  M_{\text{\RNum{2}}}=0 \, , \quad M_{\text{\RNum{3}}}=0 \, ,
\end{split}
\ee
while  the $u$-channel leads to,
\be
\ri\mathcal{A}_{\mathfrak{b}-u}=(-\ri)^3\times2e^2\int\ddl\,\rd x\,\frac{N\,l^2+O}{(l^2+m^2-x(1-x)u)^2} \, ,
\ee
\be
\begin{split}
    N_{\text{\RNum{1}}}&=0 \, , \quad N_{\text{\RNum{2}}}=\frac{-2 \ri (d-4) u }{d s} \, , \quad N_{\text{\RNum{3}}}=0 \, ,\\
    O_{\text{\RNum{1}}}&=0 \, , \quad  O_{\text{\RNum{2}}}=\frac{-2 \ri (d-4) m^2 u}{(d-2) s} \, , \quad  O_{\text{\RNum{3}}}=0 \, .
\end{split}
\ee
\subsubsection*{Type $\mathfrak{c}$ diagrams}
This type of diagrams relies on the quartic $\phi\phi^{\dagger}h_{\alpha\beta}A_{\gamma}$ interactions that arise from the following terms in the action,
\be
\begin{split}
\ri e\,\sqrt{-g}g^{\mu\nu}A_{\mu}\left(\phi\del_{\nu}\phi^{\dagger}-\del_{\nu}\phi\phi^{\dagger}\right)\supset-\ri e\, h^{\mu\nu}A_{\mu}\left(\phi\del_{\nu}\phi^{\dagger}-\del_{\nu}\phi\phi^{\dagger}\right)+\frac{1}{2}\,\ri e\,h A^{\nu}\left(\phi\del_{\nu}\phi^{\dagger}-\del_{\nu}\phi\phi^{\dagger}\right)\,.
\end{split}
\ee
This leads to the rule (momenta in-going),
\be
\mathcal{V}_{\phi\phi^{\dagger}hA}=-\ri e \eta_{\gamma(\alpha}\,(k_{1}-k_{2})_{\beta)}+\frac{\ri}{2}e\eta_{\alpha\beta}(k_1-k_2)_{\gamma}\,.
\ee
Summing the diagrams gives,
\be
\ri\mathcal{A}_{\mathfrak{c}}=2\ri^3 e^2\int\ddl\,\frac{P\,l^2}{(l^2+m^2)^2} \, ,
\ee
\be
\begin{split}
    P_{\text{\RNum{1}}}&=\frac{16\ri}{d}+0+\frac{16\ri}{d} \, , \quad P_{\text{\RNum{2}}}=0+\frac{16\ri u^2}{d s^2}+0 \, , \quad P_{\text{\RNum{3}}}=0+\frac{16\ri t^2}{d s^2}+0 \, .
\end{split}
\ee
Since the denominator of the loop integral is the same for all channels we may simply sum them up and each term in the previous $P_{\rm I,II,III}$ sums denotes the $t,s,u$-channel contribution respectively.
\subsubsection*{Type $\mathfrak{d}$ diagrams}
There is a new interaction vertex $\phi\phi^{\dagger}h_{\alpha\beta}A_{\gamma}A_{\delta}$ from the terms,
\be
-\sqrt{-g}g^{\mu\nu}e^2 A_{\mu}A_{\nu}\phi\phi^{\dagger}\supset-\frac{h}{2}\eta^{\mu\nu}e^2 A_{\mu}A_{\nu}+h^{\mu\nu}e^2 A_\mu A_\nu\,.
\ee
with rule,
\be
\mathcal{V}_{\phi \phi^\dagger h A A}=\ri e^2\left(-\eta_{\alpha\beta}\eta_{\gamma\delta}+2\eta_{\alpha(\gamma}\eta_{\delta)\beta}\right)\,.
\ee
Summing the diagrams gives,
\be
\ri\mathcal{A}_{\mathfrak{d}}=-\ri e^2\int\ddl\,\frac{Q}{(l^2+m^2)} \, ,
\ee
\be
\begin{split}
    Q_{\text{\RNum{1}}}&=-8\ri+0-8\ri \, , \quad  Q_{\text{\RNum{2}}}=0-\frac{8\ri u^2}{s^2}+0 \, , \quad   Q_{\text{\RNum{3}}}=0-\frac{8\ri t^2}{s^2}+0\,.
\end{split}
\ee
Again the terms denote the $t,s,u$-channel contributions respectively.
\subsection*{Consistency Checks}
\subsubsection*{a) Pole cancellations} The previous derivations present poles for various types of diagrams at $s=0$ and $t=0$ (for configuration \RNum{2}) and at $u=0$ (configuration \RNum{3}). In what follows we shall see that these pole contributions precisely cancel out when accounting for the wavefunction renormalisation.\\
{\bf Pole cancellation for configuration \RNum{2}:}
The sum of all $\mathfrak{a,b,c,d}$ diagrams for configuration \RNum{2} contains a pole at $s=0$ and at $t=0$ and no pole at $u=0$. The $t=0$ pole is,
\be
\mpl^2 \mathcal{A}^{\textrm{1-loop}}_{\textrm{\RNum{2}}}\supset-\frac{e^2}{2\pi^2 }\frac{u^2}{t}\int\rd x\,\rd y\,(x+y-1)^2\ln{\frac{\mu^2}{m^2}}=-\frac{e^2 }{24\pi^2}\frac{u^2}{t}\ln{\frac{\mu^2}{m^2}}\,,
\ee
directly cancelling the correction from wavefunction renormalization. The $s=0$ pole arises in the $s$-channel diagram giving the coefficient $D_{\textrm{\RNum{2}}}$ and is,
\be
\mpl^2 \mathcal{A}^{\textrm{1-loop}}_{\textrm{\RNum{2}}}\supset-\frac{e^2}{\pi^2 }\frac{u^2}{s}\int\rd x\,\rd y\,xy\ln{\frac{\mu^2}{m^2}}=-\frac{e^2 }{24\pi^2}\frac{u^2}{s}\ln{\frac{\mu^2}{m^2}}\,,
\ee
cancelling the wavefunction renormalization piece.\\
%
{\bf Pole cancellation for configuration \RNum{3}:}
The sum of all $\mathfrak{a,b,c,d}$ diagrams for configuration \RNum{3} contains a pole at $u=0$,
\be
\mpl^2\mathcal{A}^{\textrm{1-loop}}_{\textrm{\RNum{3}}}\supset-\frac{e^2}{2\pi^2 }\frac{t^2}{u}\int\rd x\,\rd y\,(x+y-1)^2\ln{\frac{\mu^2}{m^2}}=-\frac{e^2 }{24\pi^2}\frac{t^2}{u}\ln{\frac{\mu^2}{m^2}}\,,
\ee
cancelling the wavefunction renormalization piece. The $s=0$ pole arises in the $s$-channel diagram giving the coefficient $D_{\textrm{\RNum{3}}}$ and is,
\be
\mpl^2\mathcal{A}^{\textrm{1-loop}}_{\textrm{\RNum{3}}}\supset-\frac{e^2}{\pi^2 }\frac{t^2}{s}\int\rd x\,\rd y\,xy\ln{\frac{\mu^2}{m^2}}=-\frac{e^2 }{24\pi^2}\frac{u^2}{s}\ln{\frac{\mu^2}{m^2}}\,,
\ee
cancelling the wavefunction renormalization piece.
\subsubsection*{b) RG scale independence.} Again as a sanity check, we can  verify that all dependence on the renormalization scale $\mu$ drops out of the total amplitude from this gravitational exchange. We can check this explicitly for the various configurations. \\
{\bf Configuration \RNum{1}:}
Considering the type-$\fa$ diagrams we get the $\mu$ dependence,
\be
\mpl^2 \, \mathcal{A}_{\textrm{\RNum{1}}}\supset \frac{s^3+24m^2 t u}{24\pi^2 t u}e^2\ln{\frac{\mu^2}{m^2}}=\left(\frac{1}{24\pi^2}\frac{s^4}{stu}+\frac{m^2}{\pi^2}\right)e^2 \ln{\frac{\mu^2}{m^2}}\,.
\ee
The first term here cancels against the wavefunction renormalization correction to the tree amplitude. The type-$\fb$ diagrams do not give any $\mu$ dependent terms as all loop integrands are proportional to $d-4$. The remaining diagrams give an RG dependence,
\be
\mpl^2\, \mathcal{A}_{\textrm{\RNum{1}}}\supset-e^2 \frac{m^2}{\pi^2}\ln{\frac{\mu^2}{m^2}}\,,
\ee
so the total $\mu$ dependence cancels.\\
{\bf Configuration \RNum{2}:}
Considering the type-$\fa$ diagrams we get the $\mu$ dependence,
\be
\mpl^2 \, \mathcal{A}_{\textrm{\RNum{2}}}\supset \frac{su^3+12m^2 t u^2}{24\pi^2 s^2 t }e^2\ln{\frac{\mu^2}{m^2}}=\left(\frac{1}{24\pi^2}\frac{u^4}{stu}+\frac{m^2 u^2}{2\pi^2 s^2}\right)e^2 \ln{\frac{\mu^2}{m^2}}\,.
\ee
The first term here cancels against the wavefunction renormalization correction to the tree amplitude. The type-$\fb$ diagrams do not give any $\mu$ dependent terms as all loop integrands are proportional to $d-4$. The remaining diagrams give an RG dependence,
\be
\mpl^2 \, \mathcal{A}_{\textrm{\RNum{2}}}\supset-e^2 \frac{m^2 u^2}{2\pi^2 s^2}\ln{\frac{\mu^2}{m^2}}\,,
\ee
so the total $\mu$ dependence cancels.\\
{\bf Configuration \RNum{3}:}
Considering the type-$\fa$ diagrams we get the $\mu$ dependence,
\be
\mpl^2\, \mathcal{A}_{\textrm{\RNum{3}}}\supset \frac{st^3+12m^2 u t^2}{24\pi^2 s^2 u }e^2\ln{\frac{\mu^2}{m^2}}=\left(\frac{1}{24\pi^2}\frac{t^4}{stu}+\frac{m^2 t^2}{2\pi^2 s^2}\right)e^2 \ln{\frac{\mu^2}{m^2}}\,.
\ee
The first term here cancels against the wavefunction renormalization correction to the tree amplitude. The type-$\fb$ diagrams do not give any $\mu$ dependent terms as all loop integrands are proportional to $d-4$. The remaining diagrams give an RG dependence,
\be
\mpl^2\, \mathcal{A}_{\textrm{\RNum{3}}}\supset-e^2 \frac{m^2 t^2}{2\pi^2 s^2}\ln{\frac{\mu^2}{m^2}}\,,
\ee
so again the total $\mu$ dependence cancels.

\section{Spinor QED}\label{1loopspinorqed}

We now turn to the analogous derivation for spinor QED. We refer to appendix~\ref{sec:conventions} for a summary of our conventions. As mentioned in appendix~\ref{sec:Tree_level}, the tree-level photon-graviton contributions are exactly the same for scalar and spinor QED and we therefore refer to that appendix for those tree-level contributions. In what follows we can simply focus on deriving the one-loop diagrams that arise in spinor QED.

\subsection{Curved space-time action}\label{fermionscurved}
The action for spinor electrodynamics in flat space is,
\be
S=\int\rd^{4} x\left(-\frac{1}{4}F^{\mu\nu}F_{\mu\nu}+\Bar{\psi}(\ri\slashed{D}-m)\psi \right)\,.
\ee
where $\slashed{D}=\gamma^{\mu}(\del_{\mu}-\ri e A_{\mu})$. The electron is a Dirac spinor denoted $\psi$ and the Dirac adjoint is $\Bar{\psi}\equiv\psi^\dagger \gamma^{0}$ (suppressing spinor indices). The propagator for a fermion is,
\be
S_{F}=\frac{-\ri(-\slashed{p}+m)}{p^2+m^2-\ri\epsilon}\,.
\ee
To minimally couple this to gravity we use a vierbein to set up local inertial frames in which the gamma matrices take their usual constant form.
\be
S=\int\rd^{4}x\sqrt{-g} \left(-\frac{1}{4}F^{\mu\nu}F_{\mu\nu}+\Bar{\psi}\gamma^{A}v_{A}^{\phantom{A}\mu}(\ri\nabla_{\mu}+eA_{\mu})\psi-m\Bar{\psi}\psi \right)\,,
\ee
where $\nabla_{\mu}=\del_{\mu}-\frac{1}{2}\omega_{\mu AB}J^{AB}$, with $J^{AB}=\frac{1}{4}[\gamma^{A},\gamma^{B}]$ and $\omega_{\mu AB}$ is the spin connection. Here the inverse vierbein is denoted $v^{\phantom{A}\mu}_{A}$. The gamma matrices satisfy,
\be
\{\gamma^{\mu},\gamma^{\nu}\}=-2 g^{\mu\nu}\,,\quad\{\gamma^{A},\gamma^{B}\}=-2 \eta^{AB}\,.
\ee
We can write the metric as a perturbation around Minkowski space, $g_{\mu\nu}=\eta_{\mu\nu}+\kappa h_{\mu\nu}$ as well as the vierbein as a perturbation, $v_{A\mu}=\eta_{0A\mu}+\kappa c_{A\mu}$, where $\kappa=\mpl^{-1}$. As is shown in \cite{Woodard:1984sj}, the vierbein is not fundamentally necessary for the purposes of perturbation theory and can be completely eliminated in favour of the metric. Introducing the vierbein introduces six local Lorentz gauge-degrees-of-freedom which can be eliminated by imposing `Lorentz symmetric gauge',
\be
0=v_{A\alpha }\eta^{\alpha\beta}\eta_{B\beta}-v_{B\alpha}\eta^{\alpha\beta}\eta_{A\beta},
\ee
Inserting $v_{A\alpha }=\eta_{0A\alpha }+\kappa c_{A\alpha}$ we have
\ba
0&=&c_{A\alpha }\eta^{\alpha\beta}\eta_{B\beta}-c_{B\alpha}\eta^{\alpha\beta}\eta_{A\beta},  \\
&=&c_{AB}-c_{BA} \, .
\ea
This implies that $c_{A\alpha}$ is symmetric. From the definition of the vierbein one can derive,
\be
c_{\mu\nu}=\frac12 h_{\mu\nu}+\cO(\kappa)\,.
\ee
 The procedure is then to insert this gauge-fixed vierbein into the above action and expand everything to leading order in $\kappa$. The main expressions are,
\be
\begin{split}
    \sqrt{-g}&=1+\frac{\kappa h}{2}+\ldots\\
    \omega_{\mu AB}&=\frac{\kappa}{2}\left(\del_{B}h_{A\mu}-\del_{A}h_{B\mu}\right)+\ldots\\
    \Gamma^{\rho}_{\phantom{\rho}\mu\nu}&=\frac{\kappa}{2}\eta^{\rho\lambda}(h_{\mu\lambda,\nu}+h_{\nu\lambda,\mu}-h_{\mu\nu,\lambda})+\ldots\,.
\end{split}
\ee
$$
\ri\sqrt{-g}\Bar{\psi}\gamma^{A}v_{A}^{\phantom{A}\mu}\nabla_{\mu}\psi\supset\ri\frac{h\kappa}{2}\Bar{\psi}\gamma^{\mu}\del_{\mu}\psi+\ri\Bar{\psi}\gamma^{\lambda}\frac{\kappa}{2} \eta^{\mu\sigma}h_{\lambda\sigma}\del_{\mu}\psi-\ri\kappa\Bar{\psi}\gamma^{\alpha}\eta_{\alpha\phi}h^{\phi\mu}\del_{\mu}\psi+\frac{\kappa}{8}\ri\Bar{\psi}\gamma^{\mu}\del_{\alpha}h_{\beta\mu}[\gamma^{\alpha},\gamma^{\beta}]\psi
$$
$$
e\sqrt{-g}\Bar{\psi}\gamma^{A}v_{A}^{\phantom{A}\mu}A_{\mu}\psi\supset e\frac{\kappa}{2}h\Bar{\psi}\gamma^{\mu}A_{\mu}\psi+e\frac{\kappa}{2}\Bar{\psi}\gamma^{\lambda}\eta^{\mu\sigma}h_{\lambda\sigma}A_{\mu}\psi-e\kappa\Bar{\psi}\gamma^{\alpha}\eta_{\alpha\lambda}A_{\mu}\psi h^{\mu\lambda}
$$
\be
-m\sqrt{-g}\Bar{\psi}\psi\supset-m\frac{\kappa}{2}h\Bar{\psi}\psi\,.
\ee
The first operator $\ri h \kappa \Bar{\psi}\gamma^{\mu}\del_{\mu}\psi /2$ gives rise to a term in the Feynman rule (with all in-going momenta and $p_1$ being aligned with the direction of charge flow and $p_2$ being misaligned with the direction of charge flow),
\be
\ri\frac{h\kappa}{2}\Bar{\psi}\gamma^{\mu}\del_{\mu}\psi\xrightarrow[]{}\frac{\ri\kappa}{2} \eta^{\mu\nu}\gamma^{\lambda}p_{2\lambda}=\frac{\ri\kappa}{2} \eta^{\mu\nu}\slashed{p}_{2}\,,
\ee
\be
\ri\Bar{\psi}\gamma^{\lambda}\frac{\kappa}{2} \eta^{\mu\sigma}h_{\lambda\sigma}\del_{\mu}\psi\xrightarrow[]{}\frac{\ri\kappa}{2}\gamma^{(\mu}p_{2}^{\nu)}\,,
\ee
\be
-\ri\kappa\Bar{\psi}\gamma^{\alpha}\eta_{\alpha\phi}h^{\phi\mu}\del_{\mu}\psi\xrightarrow[]{}-\ri\kappa\gamma^{(\mu}p_{2}^{\nu)}\,,
\ee
\be
\frac{\kappa}{8}\ri\Bar{\psi}\gamma^{\mu}\del_{\alpha}h_{\beta\mu}[\gamma^{\alpha},\gamma^{\beta}]\psi\xrightarrow[]{}-\frac{\ri\kappa}{8}p_{3\alpha}\gamma^{(\mu}[\gamma^{\nu)},\gamma^{\alpha}]\,,
\ee
\be
-m\frac{\kappa}{2}h\Bar{\psi}\psi\xrightarrow[]{}-\ri m \frac{\kappa}{2}\eta^{\mu\nu}\,.
\ee
Summing up gives the graviton-fermion-fermion vertex,
\be
V^{\mu\nu}_{h\psi\Bar{\psi}}=\frac{\ri}{2\mpl}\left(\eta^{\mu\nu}\slashed{p}_2-\gamma^{(\mu}p_{2}^{\nu)}-\frac{1}{4}p_{3\alpha}\gamma^{(\mu}[\gamma^{\nu)},\gamma^{\alpha}]-m\eta^{\mu\nu}\right)\,,
\ee
at which point we can use the identity,
\be
\gamma^{(\mu}[\gamma^{\nu)},\gamma^{\alpha}]=\gamma^{\mu}\eta^{\alpha\nu}+\gamma^{\nu}\eta^{\alpha\mu}-2\eta^{\mu\nu}\gamma^{\alpha},
\ee
to give
\be
V^{\mu\nu}_{h\psi\Bar{\psi}}=\frac{\ri}{4\mpl}\left(\gamma^{(\mu}(p_{1}-p_2)^{\nu)}-\eta^{\mu\nu}(\slashed{p}_1-\slashed{p}_2+2m)\right)\,.
\ee
The other graviton-matter interaction vertex is,
\be
V^{\mu\nu;\alpha}_{h\psi\Bar{\psi}A}=\frac{\ri       e}{2\mpl}\left(\eta^{\mu\nu}\gamma^{\alpha}-\gamma^{(\mu}\eta^{\nu)\alpha}\right)\,.
\ee

\subsection{Photon wavefunction renormalization}

The graviton exchange diagrams are the same as in the scalar QED case, however the wavefunction renormalization factor is different due to the spinor-electron loop. As there is no photon-photon-fermion-fermion vertex, the four-point photon amplitude at one-loop involves only a subset of diagrams that we needed in the scalar QED case. The 1PI self-energy from the one diagram Fig.~\ref{fig:wfr2} gives (in $\msbar$),
\be
\Pi^{\mu\nu}=-\frac{\ri e^2}{2\pi^2}\left(\eta^{\mu\nu}k^2-k^{\mu}k^{\nu}\right)\int_{0}^{1}\rd x\,x(1-x)\ln\left(\frac{\mu^2}{m^2+x(1-x)k^2}\right)=\left(\eta^{\mu\nu}k^2-k^{\mu}k^{\nu}\right)\Pi(k^2)\,,
\ee
$$
\implies\Pi(0)=-\frac{\ri e^2}{12\pi^2}\ln{\frac{\mu^2}{m^2}}\,.
$$
As in the scalar QED case we have,
\be
Z_{A}=\frac{1}{1+\ri\Pi(0)}\,\quad\implies\quad Z_{A}=1-\frac{e^2}{12\pi^2}\ln{\frac{\mu^2}{m^2}}+\cO(e^4)\,.
\ee
Then, the one-loop corrected amplitude from the tree diagrams would be,
\be
\begin{split}
    \cA_{\textrm{tree}}^{\textrm{\RNum{1}}}&=\frac{1}{\mpl^2}\frac{s^4}{stu}\left(1-\frac{e^2}{6\pi^2}\ln{\frac{\mu^2}{m^2}}+\ldots\right)=\frac{1}{\mpl^2}\left(-\frac{s^2}{t}-\frac{s^2}{u}\right)\left(1-\frac{e^2}{6\pi^2}\ln{\frac{\mu^2}{m^2}}+\ldots\right)\\
    \cA_{\textrm{tree}}^{\textrm{\RNum{2}}}&=\frac{1}{\mpl^2}\frac{u^4}{stu}\left(1-\frac{e^2}{6\pi^2}\ln{\frac{\mu^2}{m^2}}+\ldots\right)=\frac{1}{\mpl^2}\left(-\frac{u^2}{t}-\frac{u^2}{s}\right)\left(1-\frac{e^2}{6\pi^2}\ln{\frac{\mu^2}{m^2}}+\ldots\right)\,.
\end{split}
\ee
\subsection{Non-gravitational Contributions}\label{spinornongrav}
The only one-loop non-gravitational diagrams we have are the box diagrams. The box amplitude is,
\be\begin{split}
    \ri\cA_{\textrm{box}}=(-1)(\ri e)^4\int\ddp\,\Big[&\textrm{Tr}\left(\slashed{\vep}_{1}S_{F}^{13}\slashed{\vep}_{3}S_{F}^{34}\slashed{\vep}_{4}S_{F}^{42}\slashed{\vep}_{2}S_{F}^{21}\right)+\textrm{Tr}\left(\slashed{\vep}_{1}S_{F}^{12}\slashed{\vep}_{2}S_{F}^{24}\slashed{\vep}_{4}S_{F}^{43}\slashed{\vep}_{3}S_{F}^{31}\right)\\
    +&\textrm{Tr}\left(\slashed{\vep}_{1}S_{F}^{14}\slashed{\vep}_{4}S_{F}^{42}\slashed{\vep}_{2}S_{F}^{23}\slashed{\vep}_{3}S_{F}^{31}\right)+\textrm{Tr}\left(\slashed{\vep}_{1}S_{F}^{13}\slashed{\vep}_{3}S_{F}^{32}\slashed{\vep}_{2}S_{F}^{24}\slashed{\vep}_{4}S_{F}^{41}\right)\\
    +&\textrm{Tr}\left(\slashed{\vep}_{1}S_{F}^{14}\slashed{\vep}_{4}S_{F}^{43}\slashed{\vep}_{3}S_{F}^{32}\slashed{\vep}_{2}S_{F}^{21}\right)+\textrm{Tr}\left(\slashed{\vep}_{1}S_{F}^{12}\slashed{\vep}_{2}S_{F}^{23}\slashed{\vep}_{3}S_{F}^{34}\slashed{\vep}_{4}S_{F}^{41}\right)\Big]\,,
\end{split}
\ee
where $S_{F}^{ij}$ is the Dirac fermion Feynman propagator connecting vertices with in-going photons of momenta $k_{i}$ and $k_{j}$. The factor of $(-1)$ is related to the one fermion loop. We have used the Mathematica package `Package-X' to compute the forward limit box amplitude for helicity configuration \RNum{1} and \RNum{2} \cite{Patel:2015tea}. The amplitude for configuration \RNum{3} is zero in the forward limit and the other two configurations are given by
\be
\begin{split}
    \cA^{\textrm{\RNum{1}}}_{\text{box}}(s,t=0)=\left(\frac{-e^4}{4\pi^2 s^2}\right)\Bigg\{&-4 \left(s-m^2\right) \sqrt{s \left(s+4 m^2\right)} \log \left(\frac{\sqrt{s \left(s+4 m^2\right)}+2 m^2+s}{2 m^2}\right)\\
    &-2 \left(s-2 m^2\right) \sqrt{s \left(s-4 m^2\right)} \log \left(\frac{\sqrt{s \left(s-4 m^2\right)}+2 m^2-s}{2 m^2}\right)\\
    &+\left(-4 m^4-2 m^2 s+s^2\right) \log ^2\left(\frac{\sqrt{s \left(s+4 m^2\right)}+2 m^2+s}{2 m^2}\right)\\
    &+2 m^2\left( s-2 m^2\right) \log ^2\left(\frac{\sqrt{s \left(s-4 m^2\right)}+2 m^2-s}{2 m^2}\right)+6 s^2\Bigg\}\,,
\end{split}
\ee
and
\be
\begin{split}
    \cA^{\textrm{\RNum{2}}}_{\text{box}}(s,t=0)=\left(\frac{e^4}{4 \pi ^2 s^2}\right)\Bigg\{
     &-4 \left(s+m^2\right) \sqrt{s \left(s-4 m^2\right)}  \log \left(\frac{\sqrt{s \left(s-4 m^2\right)}+2 m^2-s}{2 m^2}\right)\\
    &-2 \left(s+2 m^2\right) \sqrt{s \left(s+4 m^2\right)} \log \left(\frac{\sqrt{s \left(s+4 m^2\right)}+2 m^2+s}{2 m^2}\right)\\
    &+\left(4 m^4-2 m^2 s-s^2\right) \log ^2\left(\frac{\sqrt{s \left(s-4 m^2\right)}+2 m^2-s}{2 m^2}\right)\\
    &+2 m^2 \left(s+2 m^2\right) \log ^2\left(\frac{\sqrt{s \left(s+4 m^2\right)}+2 m^2+s}{2 m^2}\right)-6 s^2\Bigg\}\,.\hspace{-2cm}
\end{split}
\ee
So as to compare with \cite{Cheung:2014ega}, we may work below the electron mass and expanding this in powers of $s/m^2$ finally gives the same contribution for the configurations \RNum{1} and \RNum{2} to that order,
\be
\cA^{\textrm{\RNum{1}}}_{\text{box}}(s,t=0)=\frac{11 e^4 s^2}{720 \pi ^2 m^4}+\mathcal{O}(s^3/m^6)\,,\quad {\rm and}\quad
\cA^{\textrm{\RNum{2}}}_{\text{box}}(s,t=0)=\frac{11 e^4 s^2}{720 \pi ^2 m^4}+\mathcal{O}(s^3/m^6)\,.
\ee


\subsection{One-loop graviton exchange}\label{FermionQEDoneloop}
The diagrams relevant for the one-loop graviton exchange are similar to those provided in the scalar QED case (see appendix~\ref{ScalarQEDoneloop}). Referring back to Fig.~\ref{fig:scattering4}, there is no analogue to types-$\fb$ and -$\fd$ diagrams for spinor QED. In practice all type-$\fc$ diagrams vanish for spinor QED so in what follows we can simply focus our discussion on type-$\fa$ diagrams and we only provide our results for the configurations \RNum{1} and \RNum{2}.

The configurations \RNum{1} and \RNum{2} of the amplitude from type-$\fa$  diagrams are given by (expanded in the forward limit),
\be
\mpl^2 \,  \cA_{\fa}^{\textrm{\RNum{1}}}(s,t=0)=-\frac{e^2 s^2 }{6 \pi ^2 t}\ln \frac{\mu ^2}{m^2}-\frac{11 e^2 s^2}{360 \pi ^2 m^2}+\frac{e^2 s }{6 \pi ^2}\ln \frac{\mu ^2}{m^2}+\frac{e^2 s}{12 \pi ^2}\,,
\ee
and
\be
\mpl^2\, \cA_{\fa}^{\textrm{\RNum{2}}}(s,t=0)=-\frac{e^2 s^2 }{6 \pi ^2 t}\ln \frac{\mu ^2}{m^2}-\frac{11 e^2 s^2}{360 \pi ^2 m^2}-\frac{e^2 s }{6 \pi ^2}\ln \frac{\mu ^2}{m^2}-\frac{e^2 s}{12 \pi ^2}\,.
\ee
For both configurations, the first term is exactly what cancels against the one-loop wave-function renormalisation of the tree diagrams (i.e. the $t$-pole cancellation), while  the third term cancels the remaining $\mu$ dependence from the wavefunction renormalisation leading to a final amplitude that is $\mu$ independent as it should be. These results are consistent with \cite{Cheung:2014ega}. The full expressions are,
\begin{eqnarray}
\mpl^2\, \cA_{\fa}^{\textrm{\RNum{1}}}(s,t=0)&=&-\frac{e^2 s^2 }{6 \pi ^2 t}\ln \frac{\mu ^2}{m^2}+\frac{e^2}{720 \pi ^2 m^2 s} \Bigg\{120 m^2 s^2 \ln \frac{\mu ^2}{m^2}+s \left(-1560 m^4+410 m^2 s-11 s^2\right)\nn \\
&-&120 m^2 \left(s-5 m^2\right) \sqrt{s \left(4 m^2+s\right)} \ln \left(\frac{\sqrt{s \left(4 m^2+s\right)}+2 m^2+s}{2 m^2}\right)\\
&+&180 m^4\left(2 m^2- s\right) \ln ^2\left(\frac{\sqrt{s \left(4 m^2+s\right)}+2 m^2+s}{2 m^2}\right)\Bigg\}\,,\nn
\end{eqnarray}
and
\begin{eqnarray}
\mpl^2\, \cA_{\fa}^{\textrm{\RNum{2}}}(s,t=0)&=&-\frac{e^2 s^2 }{6 \pi ^2 t}\ln \frac{\mu ^2}{m^2}-\frac{e^2}{720 \pi ^2 m^2 s} \Bigg\{120 m^2 s^2 \ln \frac{\mu ^2}{m^2}+s \left(1560 m^4+410 m^2 s+11 s^2\right)\nn\\
&+&120 m^2 \left(5 m^2+s\right) \sqrt{s \left(s-4 m^2\right)} \ln \left(\frac{\sqrt{s \left(s-4 m^2\right)}+2 m^2-s}{2 m^2}\right)\\
&+&180 m^4 \left(2 m^2+s\right) \ln ^2\left(\frac{\sqrt{s \left(s-4 m^2\right)}+2 m^2-s}{2 m^2}\right)\Bigg\}\,.\nn
\end{eqnarray}


\bibliographystyle{JHEP}
\bibliography{references}

\end{document}